\documentclass[fleqn,usenatbib]{mnras}

\usepackage[T1]{fontenc}

\DeclareRobustCommand{\VAN}[3]{#2}
\let\VANthebibliography\thebibliography
\def\thebibliography{\DeclareRobustCommand{\VAN}[3]{##3}\VANthebibliography}

\usepackage{graphicx}	% Including figure files
\usepackage{amsmath}	% Advanced maths commands
\usepackage{tabularx}
\usepackage{amssymb}	% Extra maths symbols
\usepackage{orcidlink}
\usepackage{newtxtext,newtxmath}

\newcommand{\bs}{\boldsymbol}

\title[Schwarzschild model in GPU]{SchwarMAX: a GPU-friendly Schwarzschild orbit-superposition modelling framework}
\author[Zhang et al.]{%
HanYuan Zhang,$^{1}$\thanks{hz420@cam.ac.uk (hz420)}\orcidlink{0009-0005-6898-0927}
David Chemaly,$^{1}$\thanks{dc824@cam.ac.uk (dc824)}\orcidlink{0009-0001-4503-3071}
Eugene Vasiliev,$^{2}$\orcidlink{0000-0002-5038-9267}
Vasily Belokurov,$^{1}$\orcidlink{0000-0002-0038-9584}
N. Wyn Evans,$^{1}$\orcidlink{0000-0002-5981-7360}
\newauthor
Juntai Shen $^{3,4,5}$\orcidlink{0000-0001-5604-1643}
\\
$^{1}$ Institute of Astronomy, University of Cambridge, Madingley Road, Cambridge CB3 0HA, UK \\
$^{2}$ University of Surrey, Guildford GU2 7XH, UK\\
$^{3}$ Department of Astronomy, School of Physics and Astronomy, Shanghai Jiao Tong University, 800 Dongchuan Road, Shanghai 200240, People’s Republic of \\ China \\
$^{4}$ State Key Laboratory of Dark Matter Physics, School of Physics and Astronomy, Shanghai Jiao Tong University, Shanghai 200240, People’s Republic of \\  China\\
$^{5}$ Key Laboratory for Particle Astrophysics and Cosmology (MOE)/Shanghai Key Laboratory for Particle Physics and Cosmology, Shanghai 200240, \\ People’s Republic of China\\
}

\date{Accepted XXX. Received YYY; in original form ZZZ}

\pubyear{2026}

\begin{document}
\label{firstpage}
\pagerange{\pageref{firstpage}--\pageref{lastpage}}
\maketitle

\begin{abstract}
The Schwarzschild orbit-superposition method is a highly flexible dynamical modelling tool. It constrains the mass distribution of a galaxy using line-of-sight velocity and photometric observations. However, constructing such a dynamical model of a galaxy is computationally expensive. We present SchwarMAX, a new publicly available GPU implementation of the Schwarzschild orbit-superposition method. The GPU-native code is significantly faster than other implementations, with entire model construction taking around a second on GPU A100. Using SchwarMAX, we can explore the distributions of both baryonic and dark matter in a galaxy across a high-dimensional parameter space. We demonstrate its performance using mock integrated-field spectroscopic unit data generated from an N-body simulated barred galaxy. We explore the 12-dimensional space of disc, bar and halo parameters using Markov Chain Monte Carlo. %model performance against different combinations of density parameters in 12 dimensions across the dark matter halo, stellar disc, and bar parameter space. The parameter space is explored using an MCMC sampler. 
The density profiles and the bar pattern speed of the galaxy are recovered with good accuracy. We show that the code can be applied to barred galaxies across a wide range of inclination angles and can be easily extended to other stellar systems, such as elliptical and dwarf galaxies. 
\end{abstract}


%%%%%%%%%%%%%%%%%%%%%%%%%%%%%%%%%%%%%%%%%%%%%%%%%%

%%%%%%%%%%%%%%%%% BODY OF PAPER %%%%%%%%%%%%%%%%%%

\section{Introduction}

The Schwarzschild orbit-superposition method \citep{Schwarzschild1979} is one of the most flexible techniques for constructing dynamical models of galaxies. In this approach, a galaxy is represented as a weighted superposition of stellar orbits, with the aim of reproducing the observed mass and kinematic distributions while maintaining dynamical self-consistency. The distribution function is described numerically as a weighted sum of stellar orbits ($\delta$-functions in the integrals space) instead of some analytic form. This makes the Schwarzschild method particularly attractive for galaxies with complex structures, for which strong assumptions about the underlying distribution function may be too restrictive, such as barred galaxies.
%Its main idea is straightforward: for a trial gravitational potential, one integrates a large library of stellar orbits and assigns non-negative weights to them such that their superposition reproduces the observed density and kinematics of the galaxy. In this way, the stellar distribution function is represented numerically rather than through a simple analytic form. This makes the Schwarzschild method particularly attractive for galaxies with complex structures, where strong assumptions about the underlying distribution function may be too restrictive.

Another major advantage of Schwarzschild modelling over Jeans models is its ability to fit the full line-of-sight velocity distribution (LOSVD). \citet{Rix1997} introduced the use of Gauss--Hermite coefficients \citep{vanderMarel1993, Gerhard1993} to represent the LOSVD in Schwarzschild models for spherical systems. This framework was later extended to axisymmetric \citep{vanderMarel1998, Cretton1999} and triaxial systems \citep{vandenBosch2008}, leading to the \textsc{Leiden} and \textsc{Heidelberg} code families; the latter has evolved into several modern branches: \textsc{Dynamite} \citep{Jethwa2020,Thater2022} and \textsc{TriOS} \citep{Quenneville2022}. %The use of Gauss--Hermite coefficients in both orbit-based Schwarzschild models and particle-based made-to-measure (M2M) models has continued to the present day. In modern applications, Schwarzschild modelling has been generalised to systems with a wide range of geometries, with widely used codes such as \textsc{Dynamite} \citep{Jethwa2020} and \textsc{Forstand} \citep{Vasiliev2020}. 
Other implementations of the Schwarzschild method include the axisymmetric \textsc{Nukers} \citep{Gebhardt2000,Thomas2004} and \textsc{MasMod} \citep{Valluri2004} codes and triaxial \textsc{Forstand} \citep{Vasiliev2020} and \textsc{Smart} \citep{Neureiter2021} codes.
Despite differences in their detailed implementations, practical Schwarzschild models generally follow two main steps: constructing an orbital library in a trial potential and projecting it into observable space, and then combining the orbit contributions to solve for the orbital weights that best reproduce the light distribution and LOSVD of the galaxy. This procedure is repeated while exploring the model parameter space.

Over the past few decades, the Schwarzschild method has been applied to a wide range of problems in galaxy dynamics. In observational work, it has been used to constrain the masses of supermassive black holes \citep[e.g.][]{Gebhardt2000, Gebhardt2003, Saglia2016}, the stellar mass-to-light ratio \citep[e.g.][]{Krajnovic2005, Cappellari2006, vandenBosch2008}, the distribution of dark matter \citep[e.g.][]{Thomas2007}, the intrinsic shape of galaxies, and their internal orbital structure \citep[e.g.][]{Zhu2018, Jin2020}. These applications span spherical, axisymmetric, and triaxial systems, and more recently have been extended to barred galaxies as well \citep[e.g.][]{Vasiliev2020,Tahmasebzadeh2022,Tahmasebzadeh2024,Jin2025}. 

%At the level of code infrastructure, several major Schwarzschild implementations have shaped the field. Early practical observational codes include the LEIDEN family developed from spherical and axisymmetric models \citep{Rix1997,Cappellari2006} and the later triaxial HEIDELBERG implementation \citep{vandenBosch2008}. More recent packages such as \textsc{Dynamite} and \textsc{Forstand} have modernised and extended this framework \citep{Jethwa2020,Vasiliev2020}. Despite their differences, these practical observational implementations generally share the same strategy for representing the line-of-sight velocity distribution, namely through Gauss--Hermite coefficients \citep{vanderMarel1993,Rix1997}. In broad terms, Schwarzschild modelling then follows a standard sequence of steps: one specifies a trial mass model and viewing geometry, constructs and integrates an orbital library, projects the orbit contributions into the observed apertures, and solves for the orbital weights that best reproduce the luminosity and kinematic constraints. This procedure is repeated while exploring the model parameter space.

Barred galaxies are a particularly important and demanding application of this framework. Bars are common in disc galaxies \citep{Erwin2018} and play a central role in secular evolution by redistributing angular momentum and reshaping stellar and gaseous components \citep{DebattistaSellwood1998,Minchev2010}. One of their key dynamical properties is the bar pattern speed, which strongly affects resonances, orbital structure, and long-term galaxy evolution. In the past, the pattern speeds of external galaxies have usually been measured with the Tremaine--Weinberg (TW) method \citep{TremaineWeinberg1984}. 
%However, the applicability of the TW method is limited because it requires suitable inclination angles and sufficiently large spatial coverage in both the photometric and kinematic data \citep[e.g.][]{Zou2019}. Recent successes of Schwarzschild modelling for barred galaxies have also shown that these limitations can be mitigated by exploiting the full flexibility of orbit-superposition models, allowing the bar pattern speed to be constrained while simultaneously recovering the mass distribution and orbital content \citep{Tahmasebzadeh2024,Vasiliev2020, Jin2025}.
Although the TW method has its limitations, it has been applied successfully to galaxies with suitable inclination angles and sufficiently large spatial coverage in both the photometric and kinematic observations \citep[e.g.][]{Zou2019}. Recent successes of Schwarzschild modelling for barred galaxies have also shown that the bar pattern speed and the mass distribution and orbital content can be simultaneously constrained by exploiting the full flexibility of orbit-superposition models \citep{Tahmasebzadeh2024,Vasiliev2020, Jin2025}. The Schwarzschild orbital-superposition method can be applied with fewer limitations on inclination angles and data coverage. The bar of the Milky Way has already been studied with orbit-based, particle-based, and related dynamical methods \citep{Zhao1996Bar,Wang2013,Portail2017}, and explicit Schwarzschild modelling of external barred galaxies has advanced rapidly in recent years \citep{Vasiliev2020,Tahmasebzadeh2021,Tahmasebzadeh2022,Dattathri2024,Tahmasebzadeh2024,Jin2025}. These studies show that bar pattern speeds can be recovered well in controlled mock experiments, and that barred Schwarzschild models can now be applied to real barred systems.

The rise of integral-field unit (IFU) spectroscopy has made such methods increasingly valuable. Surveys such as CALIFA \citep{Sanchez2012} and MaNGA \citep{Bundy2015}, together with targeted high-resolution programmes such as TIMER \citep{Gadotti2019}, now provide spatially resolved stellar kinematics for large samples of nearby galaxies. These large data sets provide rich dynamical information on external galaxies across a wide range of morphological types, but they also amplify the main computational bottleneck of Schwarzschild modelling: for each trial potential, one must construct and integrate a large orbit library, project it into observable space, solve for the orbital weights, and repeat the procedure many times while exploring parameter space. This cost becomes especially severe for rotating barred systems, where non-axisymmetry and figure rotation must be included explicitly.

This motivates the development of a Schwarzschild framework designed from the outset for modern accelerator-based computation. In this work, we present \textsc{SchwarMAX}, a forward Schwarzschild modelling code implemented in \textsc{JAX}. \textsc{JAX} provides efficient array programming, automatic vectorisation, just-in-time compilation, and straightforward execution on GPUs, all of which are well suited to the repeated batched calculations required by Schwarzschild modelling. Rather than changing the physical basis of the method, \textsc{SchwarMAX} reimplements its most computationally expensive components in a GPU-friendly form. We validate \textsc{SchwarMAX} using mock integral-field observations generated from a barred $N$-body simulation.

The paper is organised as follows. In Section~\ref{sec:model}, we describe the setup of our model and define the likelihood function. In Section~\ref{sec:verification}, we validate the method using mock IFU observations from a barred $N$-body simulated galaxy and assess the recovery of the density profile, orbital structure, and bar pattern speed. In Section~4, we discuss the physical interpretation of the recovered constraints and the prospects for further development of the method. Finally, we summarise our results in Section~5.

\section{Model}
\label{sec:model}
The Schwarzschild method has four main parts (see Section~2 in \citealt{Vasiliev2020} for a general overview of the method): (1) constructing the gravitational potential; (2) initialising and integrating the orbits under the proposed gravitational potential; (3) collecting the line-of-sight velocity data during/after the orbital integration; (4) reweighing all the orbits to match the observed line-of-sight velocity distributions. In this section, we will describe the setup of the Schwarzschild modelling in \textsc{SchwarMAX}. 

\subsection{Density/potential construction}
\label{subsec::density}

The gravitational potential is the core of the dynamical modelling. The dark matter halo is usually represented using a parametric density model, which is later converted to potential, either through analytic density-potential pairs or with multipole expansion (usually when the model considers flattening). For the baryonic potential, many previous Schwarzschild modelling works used deprojection techniques applied to the photometric observations. The frequently adopted tools include the multi-Gaussian expansion (MGE, \citealt{Emsellem1994}) and direct fitting of the 3D parametric density profiles \citep{Dattathri2024}. Once the density profile is obtained, the potential can then be computed using azimuthal harmonic expansion (a simplified implementation of the \texttt{CylSpline} potential solver introduced in \citealt{Vasiliev2015,Vasiliev2019}), or the potential of the former MGE approach can be simply computed with a numerical 1D quadrature. These potentials are usually pre-computed before the Schwarzschild modelling at some given symmetry and inclinations. 

\textsc{SchwarMAX} has a flexible density and potential input, allowing the user to adopt all the aforementioned potential/density methods. In this work, we initialise the gravitational potential using density-potential pairs. We recommend this approach because it gives the model the highest flexibility when fitting for the photometric and kinematic data. With \textsc{SchwarMAX}'s computational speed, we can efficiently explore a high-dimensional model parameter space with 10--20 dimensions. The gravitational potential in the models presented here is composed of a dark matter halo, a stellar disc, and a rotating stellar bar.

\subsection{Orbital library construction}
\label{subsec:orbitlibrary}

With the gravitational potential specified, we construct an orbital library. In \textsc{SchwarMAX}, the orbital library is built in three stages: generation of the orbital initial conditions, orbit integration in the corotating frame of the stellar bar, and projection of the resulting orbits to the image plane.

\subsubsection{Initial conditions}  \label{sec:initial_conditions}

Traditionally, the initial conditions of orbits are assigned from a regular grid of the integrals of motions to ensure that the orbits cover sufficient volume in the phase space. For example, in an axisymmetric system, most orbits conserve three integrals of motion:  energy $E$, z-component of orbital angular momentum $L_z$, and another nonclassical integral specifying the vertical extent of the orbits, closely related to orbit inclination in a spherical system. For a non-axisymmetric system with figure rotation, the Jacobi energy, $E_J$, is the only integral of motion. To model the barred galaxies, \citet{Tahmasebzadeh2022} and \citet{Jin2025} sampled the initial condition by fixing the initial position of stars to the $x-z$ plane and sampling stars from a regular grid in $(E, \theta', R')$, where $\theta' = \arctan(x/z)$, $R' = \sqrt{x^2 + z^2}$, and $E = \Phi(x,0,z) + v_y^2/2$. The other 3 phase space coordinates, $y,~v_x,~v_z$, are kept zero for all stars. However, \citet{Vasiliev2012} showed that the initial conditions obtained from the discrete sampling method described above could cause artifacts in the resulting Schwarzschild model, and proposed an alternative scheme for sampling the initial positions and velocities randomly. Another scheme adopted in \textsc{Smart} \citep{Neureiter2021} used a regular grid in a five-dimensional starting space. %Instead, \textsc{Forstand} \citep{Vasiliev2020} used a smoother method for sampling initial conditions, and we adopt a similar approach.

%The initial spatial coordinate of the particles is sampled from a double-exponential disc with fixed scale length and scale height $p(R,z)\propto R\exp(-R/4~\rm kpc)\exp(-|z|/1.5~\rm kpc)$. This ensures that the sampled orbits cover the radial and vertical extent of the data, with more samples at the inner galaxy, where the dynamics are complicated by the presence of the galactic bar. 
Our approach for seeding the initial conditions is similar to that adopted in \textsc{Forstand}. The positions are sampled from a fiducial density distribution, namely, a fixed double-exponential profile: $p(R,z)\propto R\exp(-R/h_R)\exp(-|z|/h_z)$. This ensures that the sampled orbits cover the radial and vertical extent of the data, with more samples at the inner galaxy, where the dynamics are complicated by the presence of the galactic bar. 

The initial velocity of particles is obtained by solving the anisotropic Jeans equations in the axisymmetrised version of the barred potential. For each particle with initial spatial coordinate $\bs r$, we solve for the local velocity dispersion $\sigma_R, \sigma_\phi,~{\rm and }~\sigma_z$ and then draw the three velocity components from Gaussian distributions. The radial and vertical velocities are sampled from the Gaussian distribution centred at zero, and the Gaussian distribution of the azimuthal velocities is centred on the rotation curve. This produces a set of initial conditions for the orbits that are smoother in phase space and closer to equilibrium.

To further regularise the orbit library, each initial condition is perturbed slightly in Cartesian coordinates by sampling $n$ nearby realisations from a small uniform grid centred on its initial 6D phase-space coordinates. The size of the grid is $0.1~\rm kpc$ for the spatial coordinates and $5\%$ of the circular velocity at the location of the star, but this choice is adjustable. These $n$ realisations are treated as one orbit bundle in the later fitting. This is similar to the orbit-dithering schemes widely adopted in some other Schwarzschild studies \citep[e.g.][]{vandenBosch2008}, except that they typically create more than 20 dithers per orbit. We choose a smaller number of realisations to reduce computational cost, and this does not affect the performance of our model much, since our initial condition is intrinsically smoother than the discrete sampling method. In total, we generate $7\,500\times4$ initial conditions for later integration.

\subsubsection{Orbit integration}

The Schwarzschild orbit-superposition model requires integrating a large number of orbits ($N_{\rm orb}\approx\mathcal{O}(10^4)$) over tens to hundreds of orbital timescales. Most other implementations of the method use the eighth-order Runge--Kutta method DOP853, integrating orbits for $\sim 100-200$ dynamical times. %, similarly to \cite{Tahmasebzadeh2022} and other implementations of the method. 
The higher-order Runge--Kutta method is adopted because it provides high-accuracy interpolation along the orbits at no extra cost, allowing the orbits to be stored at regular time steps. In the present study, we adopt a simpler third-order adaptive-timestep Runge--Kutta method \citep{Bogacki1989} with a relative tolerance of $10^{-4}$. Unlike previous works, the integration is done with a fixed number of steps, rather than a target total integration time; the resulting total integration time for each orbit could vary. In practice, the orbits are typically followed for $\sim 30$--$500$ orbital times with $5\,000$ steps, which is sufficient for modelling. The fixed number of steps and the absence of branching ensures that the orbit integration can be perfectly parallelised. The Jacobi energy of the integrated orbits is typically conserved with a relative error of $10^{-4}$--$10^{-2}$. To remove bad orbits from our model, orbits with a relative Jacobi energy drift larger than $10^{-1}$ are discarded from the library, but only less than $0.1\%$ of the orbits will be discarded from this quality cut.

The orbit integration is performed in the reference frame of the galactic bar rotating with a pattern speed $\Omega_p$, which is also a free parameter in our model. The major axis of the galactic bar is always aligned with the x-axis during the orbit integration. The velocities in the bar rest frame are transformed back to the inertial frame after the orbital integration for later comparison with the observation. 

Because figure rotation breaks the full eightfold symmetry available in stationary triaxial models, only a fourfold symmetry remains for barred systems \citep{Deibel2011, Vasiliev2020, Tahmasebzadeh2022}. We therefore generate three mirror copies of each integrated orbit by flipping the signs of positions and velocities, as summarised in Table~\ref{tab:orbit_symmetry}. These mirrored trajectories are co-added and treated as a single symmetrised orbit bundle during fitting. This increases the effective orbit sampling at negligible additional cost and enforces the assumed symmetry of the rotating barred potential.

\begin{table}
\centering
\caption{Fourfold symmetrisation applied to each integrated orbit in the barred model.}
\label{tab:orbit_symmetry}
\begin{tabular}{lll}
\hline
Copy & Position & Velocity \\
\hline
1 & $(x,\,y,\,z)$ & $(v_x,\,v_y,\,v_z)$ \\
2 & $(x,\,y,\,-z)$ & $(v_x,\,v_y,\,-v_z)$ \\
3 & $(-x,\,-y,\,z)$ & $(-v_x,\,-v_y,\,v_z)$ \\
4 & $(-x,\,-y,\,-z)$ & $(-v_x,\,-v_y,\,-v_z)$ \\
\hline
\end{tabular}
\end{table}

\subsubsection{Projection to the image plane}
\label{subsubsection:projection}
After orbit integration, the library must be transformed from the intrinsic bar frame to the observer's frame before comparison with IFU data. In \textsc{SchwarMAX}, this projection is performed directly on the orbit library.

We define the intrinsic coordinates $(x,y,z)$ such that the $x$-axis is aligned with the bar major axis, the $y$-axis lies in the disc plane, and the $z$-axis is perpendicular to the disc. We then rotate the orbit library to the observer's frame using the $z$-$x$-$z$ Euler-angle convention. The rotated coordinates $(x',y',z')$ are defined by
\begin{equation}
\begin{pmatrix}
x'\\ y'\\ z'
\end{pmatrix}
=
\mathbf{R}(\alpha,\beta,\gamma)
\begin{pmatrix}
x\\ y\\ z
\end{pmatrix},
\qquad
\mathbf{R}(\alpha,\beta,\gamma)
=
\mathbf{R}_z(\gamma)\,\mathbf{R}_x(\beta)\,\mathbf{R}_z(\alpha),
\end{equation}
where $\alpha,~\beta,~\gamma$ are the Euler's angles, and $\mathbf{R}_i$ is the rotation matrix along $i$-axis. The three Euler angles are also the free parameters in our model. 
After this convention, we define the image plane as given by $(X,Y)=(x',y')$, while the line of sight is along the $z'$ direction. The projected line-of-sight velocity is therefore taken to be the $z'$ component of the velocity after rotation. In this projection scheme, the position angle of the galaxy is controlled by $\gamma$, inclination is controlled by $\beta$, and the in-plane angle of the galactic bar is controlled by $\alpha$.
%`This choice is convenient for the present implementation, because after a single Euler rotation, the projected positions and velocities can be accumulated directly on the sky plane.

\subsubsection{Mapping orbital library to observations}

After orbit integration and projection, we convert the orbital library into the observables used in the fit later. Because the adaptive integrator outputs samples at irregular time intervals, each sample along the orbit is reweighted linearly by its corresponding integration time step size $\delta t$. The weight of each sample along the orbit is $\delta t/t_{\rm total}$, where $t_{\rm total}$ is the total integration time of the orbit.

For the self-consistency constraints, we divide the 3D cylindrical space into $N_{\rm dens}$ bins uniformly spaced in $(R,\,\phi,\,z)$ direction that cover the radial and vertical extents of the data. In each orbit, we compute its mass contribution to each of the 3D cells and record this as a matrix ${\bf m}_{i,n}$, where $i$ denotes the orbit index and $n$ denotes the bin index. In parallel, we also compute the targeting enclosed stellar mass in each spatial cell from the density/potential used for the orbit integration ${\bf m}^*_{n}$. This is later used to optimise the orbital weight, ensuring that the stars in the resulting models are self-gravitating.

The surface luminosity density and line-of-sight velocity distribution (LOSVD) are usually considered observational constraints on the Schwarzschild model. We, therefore, need to compute the corresponding statistics from the orbital library. The galaxy projected on the 2D image plane is usually discretised into a large number of 2D apertures, using methods such as Voronoi binning \citep[][]{Cappellari2003}, where the binning strategy is pre-defined before modelling using \textsc{PowerBin} \citep{Cappellari2025}. Similarly to the self-consistency constraints mentioned before, we compute the surface density contributed by each orbit $i$ for each 2D aperture $n$ as a matrix. The orbits are collected first in regular bins and then regrouped into Voronoi bins according to the pre-defined binning strategy. We then use a mass-to-light ratio $M/L \equiv \Upsilon$, to convert the surface mass density to the surface luminosity density, which we denote as ${\bf \Sigma}_{i,n}$. The $M/L$ ratio is constant across the galaxy, making it another free parameter in the model. Previous studies have explored spatially varying $M/L$ ratio \citep{McConnell2013, Erwin2018, Mehrgan2024}. While this is also feasible in our modelling framework, it would significantly complicate the parameterisation, so we do not explore it in the model presented here. We note that most other implementations of the Schwarzschild method employ simultaneous rescaling of mass by $\Upsilon$ and velocity by $\sqrt{\Upsilon}$ to reuse the same orbit library for different choices of the $M/L$ ratio, as explained in section 2.3 of \cite{Vasiliev2020}. This scales all mass components (not just stars) in the same proportion. In the present code, however, the orbit library is constructed afresh for every new choice of model parameters, and the $M/L$ factor is applied to the stellar component only, to simplify the interpretation. % It is also noteworthy that previous pipeline conventionally rescale the masses of all galaxy mass components by the $M/L$ ratio, which could save the effort of orbital library construction as one could simply rescale the velocity axis when varying the $M/L$ ratio \citep[e.g.][]{Vasiliev2020, Jethwa2020}. However, as we use later use MCMC to explore the parameter space, it is non-trivial to apply this symmetry into our pipeline.

The Gauss--Hermite polynomial moments are usually used to describe the LOSVD in each 2D aperture \citep[][]{Gerhard1993, vanderMarel1993}. The Gauss--Hermite coefficients, $h_m$, are defined through the Gauss--Hermite expansion:
\begin{equation}
    f(v_{\rm los}) = \frac{A}{\sqrt{2\pi}\,s}\exp\left(-\frac{w^2}{2}\right)\left[1 + \sum_{m=1}^{N_{\rm GH}} h_m {\mathcal H}_m(w)\right],
\end{equation}
where $w=(v_{\rm los}-v_0)/s$, $v_0$ and $s$ are the pre-defined centre and width of the reference Gaussian, ${\mathcal H}_m$ are the modified Hermite polynomials, $A$ is a normalisation constant, and $N_{\rm GH}$ is the highest order of the adopted Gauss--Hermite coefficients. Due to the linearity of the Gauss--Hermite coefficients, we compute $h_m$ of every orbit $i$ in every 2D aperture $n$ and then sum among the orbits after applying an orbital weighting. Therefore, we represent the LOSVD of the orbital library as four matrices ${\bf h}_{m,i,n}$, where $m = \{1,2,3,4\,...\}$ stands for the orders of the GH coefficient.

Overall, we restore the orbital library into $N_{\rm GH}+2$ matrices encoding the mass, and LOSVD contributed by each orbit to each 2D or 3D bin, which we will later use to compute the orbital weights and the goodness-of-fit of our model. 

\subsection{Optimisation of orbital weights}

For a fixed gravitational potential, the remaining freedom in the Schwarzschild model lies in the non-negative orbital weights, which determine how strongly each orbit contributes to the intrinsic 3D density, projected surface density, and LOSVD. We therefore reweight the orbital library to obtain the best representation of the observed density and kinematics under the specified potential. In the present implementation, the orbital weights, $w_i$, are determined by minimising the objective function
\begin{equation}
\begin{aligned}
Q = \frac{1}{2}\Bigg[
&\overbrace{
\sum_{n=1}^{N_{\rm dens}}
\left(\frac{\sum_{i=1}^{N_{\rm orb}} w_i m_{i,n}-m^*_n}
{0.01\,m^*_n}\right)^2
}^{\chi^2_{\mathrm{dens}}} +
\overbrace{
\sum_{n=1}^{N_{\rm obs}}
\left(\frac{\sum_{i=1}^{N_{\rm orb}} w_i \Sigma_{i,n}-\Sigma^*_n}
{0.01\,\Sigma^*_n}\right)^2
}^{\chi^2_{\rm lum}}
\\[1pt]
&+
\underbrace{
\sum_{m=1}^{N_{\rm GH}}\sum_{n=1}^{N_{\rm obs}}
\left(\frac{\sum_{i=1}^{N_{\rm orb}} w_i h_{m,i,n}-h^*_{m,n}}
{\Delta h^*_{m,n}}\right)^2
}_{\chi^2_{\mathrm{kine}}}
\Bigg] +
\underbrace{
\frac{\lambda}{N_{\mathrm{orb}}}
\sum_{i=1}^{N_{\mathrm{orb}}} \left(\frac{w_i}{\widetilde{w_i}}\right)^2
}_{\text{regularisation}}
\end{aligned}
\label{eqn:objective_function}
\end{equation}
subject to $w_i \ge 0$, where $\Sigma^*$ is the observed surface luminosity density, $h^*_{m}$ is the observed m-th Gauss--Hermite coefficients, $\Delta h^*_{m}$ is the corresponding uncertainty, and $\widetilde{w_i}$ is the orbital weights if all orbits contribute equally. The objective function is a summation of the $\chi^2$ of the self-consistency constraints, surface luminosity density constraints, kinematics constraints, and a regularisation term. Following \citet{Vasiliev2020}, we adopt a diagonal quadratic regularisation term to smooth the orbital weight space by encouraging equal orbital weights. The tolerance of the self-consistency density constraints and surface luminosity constraint is set at the level of $1\%$, the same as the default setting in \textsc{Dynamite} \citep[e.g.][]{Tahmasebzadeh2022}.

Eqn.~\ref{eqn:objective_function} can be re-written in the compact form
\begin{equation}
Q(\mathbf{w})=
\frac{1}{2}\left\lVert \mathbf{U}\mathbf{w}-\mathbf{y}\right\rVert^2
+\frac{\lambda}{N_{\mathrm{orb}}}\left\lVert \mathbf{w}\right\rVert^2,
\qquad
\mathbf{w}\ge 0,
\end{equation}
where $\mathbf{U}$ is constructed by stacking all the matrices ($m_{i,n},~\Sigma_{i,n},~h_{m,i,n}$) obtained before, and ${\bf y}$ is the corresponding stacked targets. This optimisation problem fits naturally into a non-negative least-squares (NNLS) formulation, which has also been adopted in earlier Schwarzschild studies \citep[e.g.][]{vandenBosch2008}. The alternative used in \cite{Vasiliev2020} is a more general quadratic programming (QP) problem, which permits both equality constraints (for the 3d density, i.e.\ gravitational self-consistency) and least-square fitting of remaining kinematic constraints. We solve this NNLS problem using the Alternating Direction Method of Multipliers (ADMM). This approach is particularly well suited to the present application because the objective function is quadratic and convex. While we treat the self-consistency constraints as a part of the objective function, the only explicit constraint is the non-negativity of the orbital weights. In each ADMM iteration, the NNLS problem is split into two parts: solving the unconstrained quadratic problem with a linear solver and enforcing the non-negative constraints by clipping negative orbit weights to zero. A vector that tracks the differences between the solution before and after the constraint is recorded and reused at the next iteration to accelerate the convergence. 

\citet{Magorrian2006} highlighted the importance of marginalising over all possible distribution functions allowed by each potential when comparing the goodness-of-fit of different potentials. \citet{Bovy2018} applied this idea to a made-to-measure model. For our Schwarzschild modelling with high dimensionality, it is still too computationally expensive to evaluate this faithfully, even when implemented on a GPU. Here, we try to account for different possible distribution functions approximately by exploring the underlying uncertainties on the best-fit orbital weights. To do so, we generate 100 bootstrap realisations of the observed surface luminosity density map and the GH coefficient maps by perturbing them according to their measurement uncertainties. The intrinsic three-dimensional density constraint is not bootstrapped, since it is fixed by the specified potential-density model. For each bootstrap, we solve the same NNLS problem and obtain one best-fit set of orbital weights. Because all realisations share the same orbital library and the same system matrix $\mathbf{U}$, only the data vector $\mathbf{y}$ changes between realisations, allowing the 100 optimisations to be carried out in parallel. This is especially convenient in the present JAX/GPU implementation as it could be achieved without significant increase in the computational time. The resulting ensemble sets of orbital weights are then used to marginalise over orbital weight distributions in the final likelihood evaluation later. Although this is motivated by \citet{Magorrian2006}, our approach is still different from what was proposed in \citet{Magorrian2006} to marginalise over all possible distribution functions. The bootstrap we performed here only samples the orbital weights distribution given the potential and data. To marginalise over orbital weight space, it would require us to sample large numbers of orbital weights that satisfy the self-consistency and non-negative constraints. This is still unfeasible with our approach. The marginalisation of the bootstrapped orbital weight is more similar to the method of measuring orbital flexibility as proposed in \citet{Lipka2021}, in which they penalise models with large likelihood variation during the bootstrap. 

\subsection{Model likelihood}
\label{subsec::model_likelihood}
\begin{figure*}
    \centering
    \includegraphics[width= \textwidth]{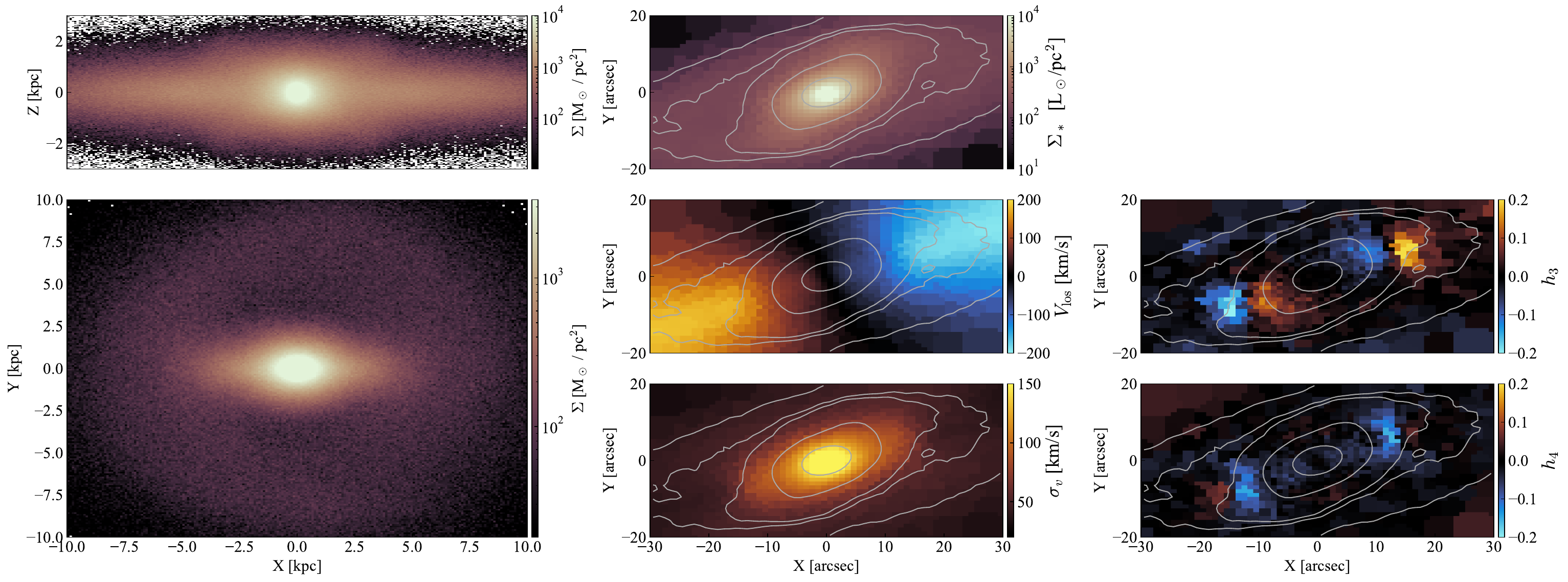}
    \caption{A summary of the mock IFU dataset we used to validate the method. {\it Left:} the edge-on and face-on view of the N-body simulated galaxy for the mock data generation. {\it Middle:} the surface luminosity density map ($\Sigma_{*}$), the mean line-of-sight velocity map ($V_{\rm los}$), and the line-of-sight velocity dispersion map ($\sigma_v$) after Voronoi binning with projection Euler angles $(\alpha, \beta, \gamma) = (45^\circ, 65^\circ, 25^\circ)$. The contour denotes the surface density of the galaxy in the image plane. {\it Right:} the $h_3$ and $h_4$ maps.}
    \label{fig:mock_data_summary}
\end{figure*}

Once the orbital library has been constructed and the orbital weights have been determined, we evaluate the final model likelihood for a given set of density parameters. In the present implementation, the intrinsic three-dimensional density enters as a self-consistency constraint only in the orbital weight optimisation, while the final observational likelihood is defined from the projected surface luminosity density map and the kinematic maps only (i.e., excludes $\chi^2_{\mathrm{dens}}$). For the $b$-th bootstrap realisation of the orbital weights, we write the model $\chi^2$ as
\begin{equation}
\chi^2_b = \chi^2_{{\rm lum},b} + \chi^2_{{\rm kine},b},
\end{equation}

To account approximately for the uncertainty induced by the orbital-weight determination, we combine and marginalise the individual $\chi^2$ from the $B=100$ bootstrap realisations through,
\begin{equation}
\chi_{\rm model}^2 ({\bs \theta})
=
\ln\left[
\sum_{b=1}^{B}
\exp\left(\frac{\chi^2_{b, \rm min}-\chi^2_b}{2}\right)
\right] + \rm const,
\end{equation}
where $\chi^2_{b, \rm min}$ is the minimum $\chi^2_b$ among all the bootstraps. $\chi_{\rm model}^2 ({\bs \theta})$ is the final $\chi^2$ of the model given model parameters $\bs \theta$. This quantity serves as the effective likelihood for a given set of density parameters after approximately marginalising over the uncertainty in the orbital weights optimisation.

Before sampling the posterior distribution, we first launch a minimiser to identify the density parameters that minimise $\chi_{\rm model}^2$ and thus provide the best-fit solution. We then estimate the uncertainty scale of the resulting $\chi_{\rm model}^2$ using a jackknife procedure over the observational apertures, namely, repeat the NNLS solution $N_{\rm aper}$ times, excluding the data for one aperture at a time, and compute the scatter in the resulting $\chi^2$ values. % \hanyuan{(masking a 2D aperture out per time of NNLS)}, similar in spirit to previous implementations of the Schwarzschild method \citep[e.g.][]{Zhu2018, Tahmasebzadeh2022, Jin2025}. 
This provides an empirical estimate of the model uncertainty represented in $\Delta \chi^2$, which is used to rescale the likelihood surface in the subsequent parameter inference. The final log-likelihood of the model is defined as 
\begin{equation}
    \ln \mathcal{L} = -\chi_{\rm model}^2 / \Delta \chi^2
\end{equation}
The use of $\Delta \chi^2$ as the denominator of the final model likelihood is an empirical approach to smooth the log-likelihood in the parameter space, dating back to \citet{vandenBosch2008} and adopted in many subsequent studies \citep[e.g.][]{Zhu2018, Tahmasebzadeh2022, Jin2025}. However, it lacks a rigorous statistical foundation, and its primary motivation is to avoid extremely narrow uncertainties on model parameters when using the standard criterion ($\chi_{\rm model}^2 - \chi_{\rm model,\, min}^2 = 1, 4, 9$, etc., for 1, 2, 3-$\sigma$ intervals). A more statistically motivated approach would require marginalisation over the orbital weights space as emphasised in \citet{Magorrian2006}, which is still computationally infeasible so far. However, our work could present a feasible pathway towards this goal (see discussion in Section~\ref{subsec::future_develop}).

Another, more subtle reason for rescaling the log-likelihood values is the intrinsic stochasticity of the $\chi^2$ values computed from a finite-size orbit library. One source of it is the random assignment of orbital initial conditions in each new model (Section~\ref{sec:initial_conditions}). However, even if one uses the same orbital initial conditions in all trial potentials (which may not be optimal if the depth of the potential well, and hence the characteristic velocity of orbits, varies significantly across the model parameter space), there remains another source of fluctuations in $\chi^2$ especially relevant for barred galactic potentials, which have a significant fraction of chaotic orbits. These trajectories have rather unpredictable variations even for very small differences in the potential, which translates into the noise in the matrices ${\bf m}_{i,n}$ describing contribution of orbits to constraints, and finally into the noise in the objective function (Equation~\ref{eqn:objective_function}). Although it might be tempting to banish those chaotic orbits from the model entirely, they are important building blocks of the galaxy especially around the corotation region \citep[e.g.][]{Voglis2007,Manos2011}, and models without them might not be feasible at all \citep{Merritt1996,Vasiliev2012}, although this topic has not been explored specifically for barred galaxies. In any case, our experiments suggest that merely varying the ODE integration tolerance can lead to changes in $\chi^2$ at the level $\mathcal O(10)$, so a rescaling factor $\Delta\chi^2 \gg 1$ seems necessary.

\section{Verification}
\label{sec:verification}

\subsection{N-body simulation and mock IFU observation}

We validate our model using a mock IFU data set constructed from a barred galaxy. As the source model, we adopt the Milky Way analogue $N$-body simulation of \citet{TG21}, which consists of a stellar bulge and disc with masses of $1.3\times10^{10}~M_\odot$ and $4.3\times10^{10}~M_\odot$, respectively, embedded in a dark matter halo of mass $1.2\times10^{12}~M_\odot$. The galaxy has a Milky Way-like rotation curve. A stellar bar forms naturally $\sim 1~\rm Gyr$ after the start of the simulation, and vertical buckling occurs at around $\sim 2~\rm Gyr$. We select two snapshots, at $4$ and $7$~Gyr after the start of the simulation, to generate mock observations at different projection angles. The bar pattern speeds of these two snapshots are $33~\rm km/s/kpc$ and $25~\rm km/s/kpc$, respectively. In this paper, we mainly present the results for the snapshot at $7$~Gyr. The face-on and edge-on views of this snapshot are shown in the left panels of Fig.~\ref{fig:mock_data_summary}.

We place the galaxy at a fiducial distance of $50~\rm Mpc$, for which $1'' \approx 0.25~\rm kpc$. We then project the galaxy onto the image plane following the same procedure described in Section~\ref{subsubsection:projection}, using Euler angles $\alpha=45^\circ$, $\beta=65^\circ$, and $\gamma=25^\circ$. This corresponds to an inclination of $65^\circ$, with the bar major axis viewed approximately $45^\circ$ on in the image plane. We generate the mock IFU data within a field of view of $60''\times40''$ (typical of IFU instruments such as SAURON or MUSE) centred on the galaxy. The pixels are then grouped using Voronoi binning \citep{Cappellari2003} with the recently updated package \textsc{PowerBin} \citep{Cappellari2025}, such that the rebinned apertures have approximately uniform flux. After binning, $\approx600$ apertures remain. For each aperture, we calculate the mean and standard deviation of the LOSVD, $V_{\rm los}$ and $\sigma_v$, and fix $v_0$ and $s$ when computing the Gauss--Hermite coefficients such that $h_1 = h_2=0$. We assign kinematic uncertainties typical of modern observations, namely $\Delta V_{\rm los}, \Delta \sigma_{\rm v}=5~\rm km/s$ and $\Delta h_{3,4}=0.03$. The uncertainties on $h_1$ and $h_2$ are propagated as $\Delta h_1 = \Delta V_{\rm los}/\sqrt{2}s$ and $\Delta h_2 = \Delta \sigma_{v}/\sqrt{2}s$ \citep{Rix1997}. The surface mass density is converted to a surface luminosity density assuming a constant mass-to-light ratio of $1$. The resulting mock observations are shown in the middle and right columns of Fig.~\ref{fig:mock_data_summary}.

\subsection{Model construction}  \label{sec:mock_test_model_setup}

\begin{figure*}
    \centering
    \includegraphics[width= \textwidth]{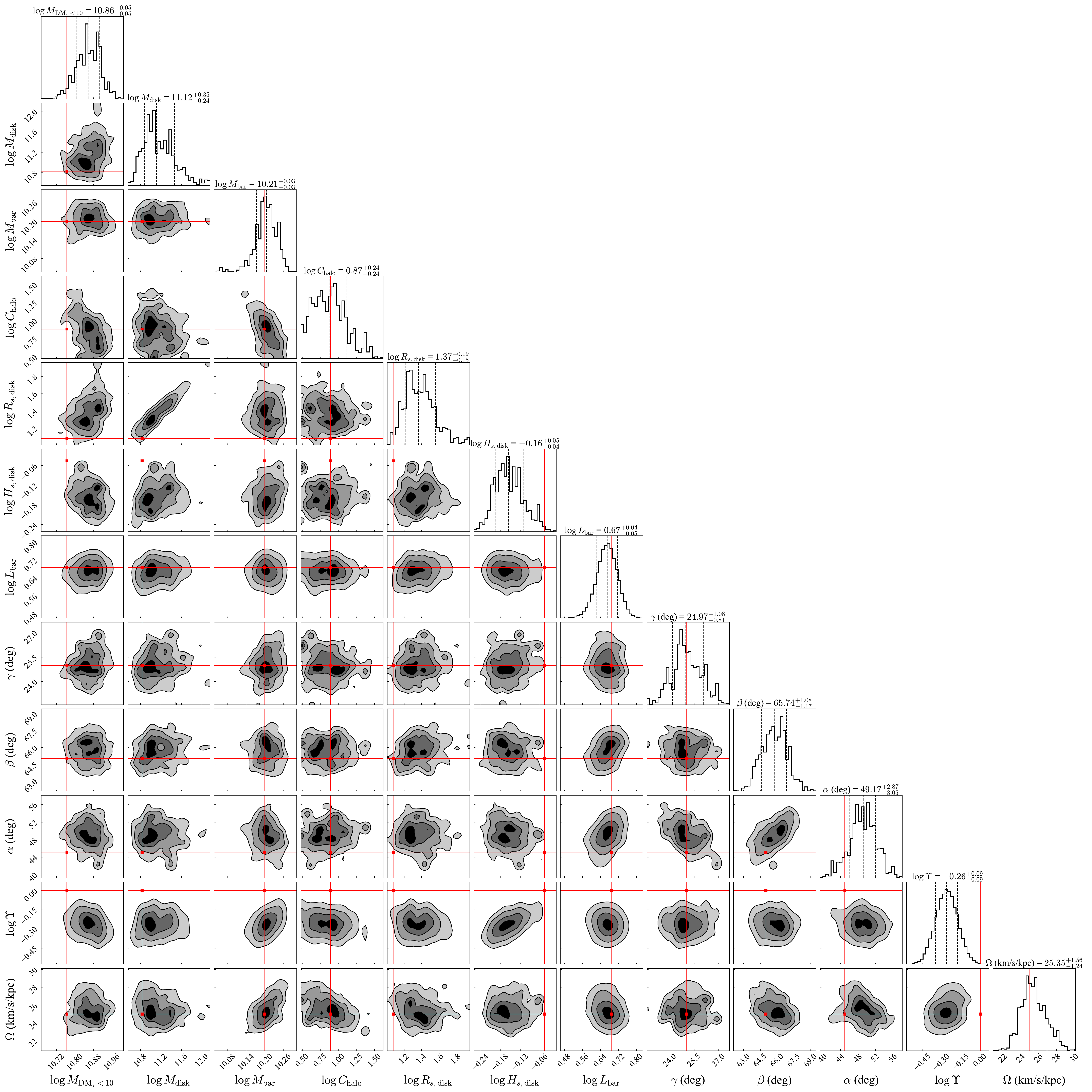}
    \caption{The corner plot of the posterior of the model parameters fitted to the mock generated data. The ground truth value for each parameter are denoted in red lines for reference. The ground truth are obtained by fitting the 3D density distribution of dark matter and stars. }
    \label{fig:posterior_main}
\end{figure*}

\begin{figure*}
    \includegraphics[width=\textwidth]{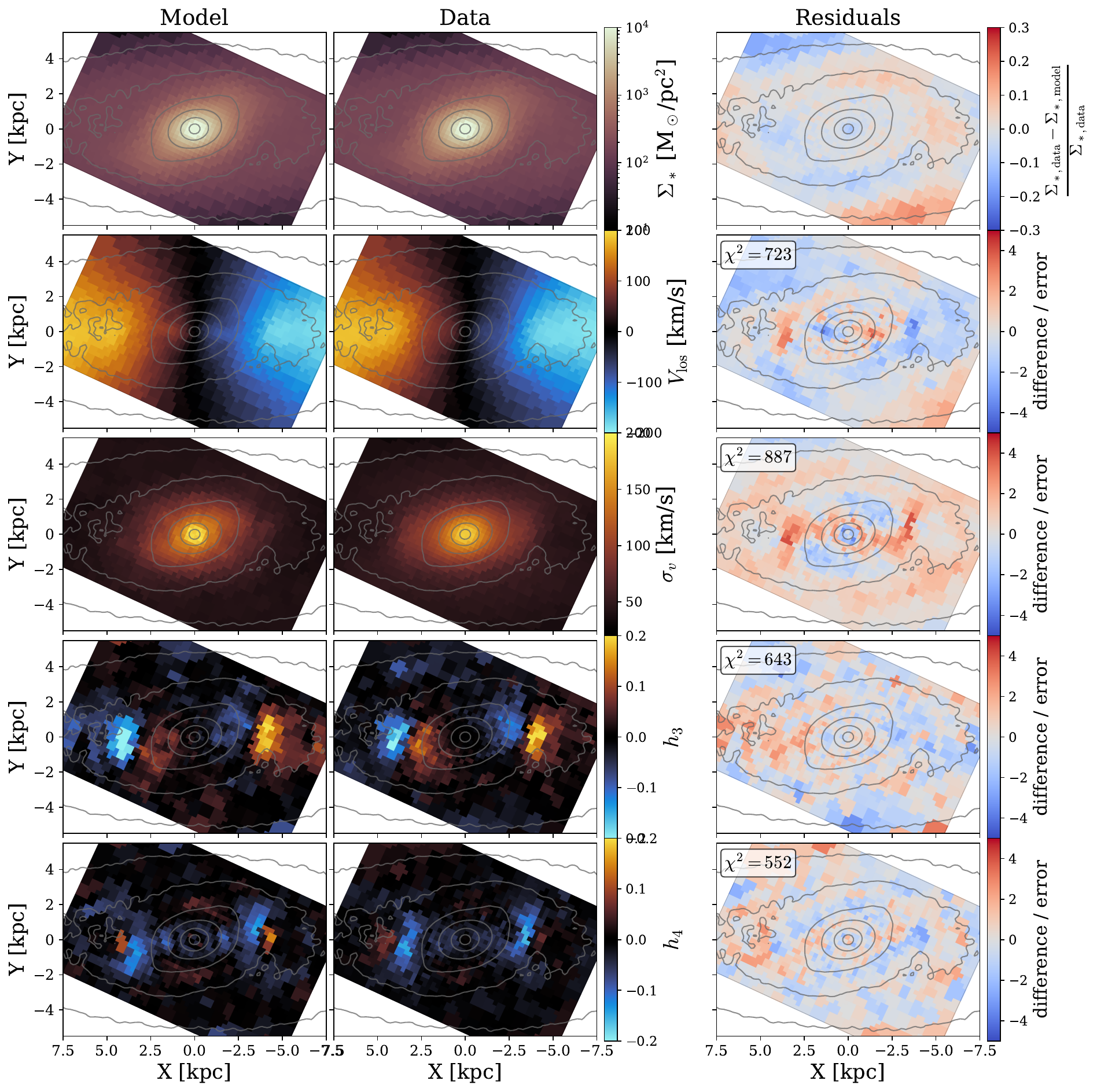}
    \caption{A comparison between the mock data and the best-fit model. The five rows from top to bottom show the surface luminosity density, $V_{\rm los}$, $\sigma_v$, $h_3$, and $h_4$ maps, respectively. The left column show the best-fit model, the middle column show the mock data, and the right panel shows the residual. The residual of the surface density are shown as a fraction of the mock observed surface density, while the residual of the other kinematic maps are scaled by the assigned uncertainty.}
    \label{fig:data_vs_model}
\end{figure*}

We describe the model setup in order to fit the mock IFU observation of this barred galaxy, including the choice of the gravitational potential and model parameters sampling method.

\subsubsection{Dark matter halo}
\label{subsubsec::nfw}

We represent the dark matter halo with the spherical NFW profile \citep{NFW1996}:
\begin{equation}
\begin{aligned}
        \rho_{\rm NFW}(r) &= \frac{M_{\rm DM}}{4\pi R_s^3}\frac{1}{\dfrac{r}{R_s}\left(1 + \dfrac{r}{R_s}\right)^2},\\
        \Phi_{\rm NFW}(r) &= -\frac{GM_{\rm DM}}{r}\ln\left(1 + \frac{r}{R_s}\right)
\end{aligned}
\end{equation}
where $M_{\rm DM}$ is the characteristic mass parameter and $R_s$ is the scale radius; both are free parameters in our model. Because the kinematic data mainly constrain the mass distribution within the observed radial extent, a degeneracy between $M_{\rm DM}$ and $R_s$ is expected. To alleviate this degeneracy, we re-parameterise the halo in terms of the enclosed dark matter mass within a reference radius $R$, $M_{\rm DM}(<R)$, and the halo concentration $c$, where
\begin{equation}
\begin{aligned}
 & M_{\rm DM}(<R) = 4\pi \rho_0 R_s^3
\left[
\ln\!\left(\frac{R_s + R}{R_s}\right)
- \frac{R}{R_s + R}
\right], \\
& c = \frac{R_{\rm vir}}{R_{s}},
\end{aligned}
\end{equation}
and
$R_{\rm vir} = \left( 3M_{\rm vir}/4\pi\,\Delta_\mathrm{crit}\, \rho_{\mathrm{crit}} \right)^{1/3}$, with $M_{\rm vir}$ the virial mass and $\Delta_\mathrm{crit}=200$ the standard choice of critical overdensity. %[CITATION NEEDED].

\subsubsection{Stellar disc and bar}

The stellar disc is represented by a Miyamoto-Nagai disc \citep[][]{Miyamoto_Nagai1975}, for which the gravitational potential is
\begin{equation}
    \Phi_{\rm MN}(R, z) = - \frac{GM_{\rm d}}{\mathcal R},
\end{equation}
where $\mathcal R^2=R^2+(R_d+\zeta)^2$ and $\zeta=\sqrt{z^2+H_{d}^2}$. The corresponding density profile is
\begin{equation}
    \rho_{\rm MN}(R, z) = \left( \frac{H_{d}^2 M_{\rm d}}{4\pi} \right)
\frac{
R_d \mathcal{R}^2 + 3\zeta(\zeta+R_d)^2}{\mathcal R^{5} \zeta^{3}},
\end{equation}
where $M_{\rm d}$ is the total mass of the stellar disc, and $R_d$ and $H_d$ are the radial scale length and vertical scale height, respectively. These three quantities are free parameters in our model.

% For the stellar bar, we adopt density-potential pairs recently developed in \citet{Dehnen_Aly2023}. They derived a family of density-potential pairs by convolving the Miyamoto-Nagai disc profile with a needle function, similar to the approach in \citet{Long_Murali1992}.

For the stellar bar, we adopt the analytical bar models of \citet{Dehnen_Aly2023}. These models are constructed by convolving an axisymmetric parent potential-density pair with a finite one-dimensional needle density along the bar major axis, following the general approach of \citet{Long_Murali1992}. The parent potential-density pairs form two related families of axisymmetric disc models built from higher-order modifications of the Miyamoto-Nagai disc, which they dubbed $T$ and $V$ potential families. For a brief summary of their approach, if the parent model contains terms of the form \(\mathcal R^{-n}\), then the barred model is obtained through the replacement
\begin{equation}
\mathcal R^{-n}\rightarrow I_n(x,y,Z;L,\eta)
\equiv
\int_{x-L}^{x+L}
\frac{f(x - x')\,dx'}{\left[x'^2+y^2+Z^2\right]^{n/2}},
\label{eqn:Rn_to_In}
\end{equation}
where \(L\) is the bar half-length, \(Z = a + \zeta\), and
\[
f(x)=
\frac{1}{2L}\left[1+\eta\left(1-\frac{2|x|}{L}\right)\right]
\]
for $|x|<L$, and zero otherwise. Here \(\eta\), with \(-1\leq\eta\leq1\), controls the mass gradient along the bar major axis: larger \(\eta\) implies a more centrally concentrated distribution along the bar.

In our implementation, the stellar bar is represented by a long-bar component and a boxy/peanut bulge component, modelled with the $T_3$ and $V_4$ potential of \citet{Dehnen_Aly2023}, respectively. The analytic expression of these two potentials is in Appendix~\ref{Appendix:Dehnen_ALy_bar}. In the general \citet{Dehnen_Aly2023} barred model, each component has five parameters: the mass $M_b$, the in-plane scale length $a$, the vertical scale height $b$, the needle half-length $L$, and the needle-slope parameter $\eta$. The parameter $a$ sets the radial scale of the parent axisymmetric model, $b$ sets the vertical thickness, $L$ approximates the bar half-length, and $\eta$ controls the mass gradient along the bar major axis.

To keep the number of free parameters to a minimum, we tie the masses of the $T_3$ and $V_4$ components and denote their common normalisation by $M_b$. We further fix $\eta = 1$ for $T_3$ and $\eta=0$ for $V_4$, adopt $(a,~b,~L)_{V_4} = (0.5, 0.5, 0.1)~\rm kpc$, and set $(a,~b,~L)_{T_3} = (L_b/5, b_{\rm d}, L_b)$. The stellar bar model therefore has two free parameters, $M_b$ and $L_b$, which represent the total bar mass and the bar half-length.

\begin{table*}
\centering
\caption{Summary of the free parameters in our model, $12$ in total}
\label{tab:model_parameters}
\begin{tabular}{lll}
\hline
Parameter & Meaning & Sampling prior \\
\hline
$\log M_{\rm DM}(<10~\rm kpc)$ & enclosed dark-matter mass within $10\,{\rm kpc}$ & $\mathcal{U}(8.5,14.5)$ \\
$\log M_{\rm d}$ & stellar disc mass & $\mathcal{U}(7.5,13.5)$ \\
$\log M_{\rm b}$ & stellar bar mass & $\mathcal{U}(7.5,13.5)$ \\
$\log c$ & halo concentration parameter & $\mathcal{U}(0,2)$ \\
$\log R_{\rm d}$ & disc radial scale length & $\mathcal{U}(-0.5,1.5)$ \\
$\log H_{\rm d}$ & disc vertical scale height & $\mathcal{U}(-1.5,0.5)$ \\
$\log L_{\rm b}$ & bar half-length & $\mathcal{U}(-0.5,1.5)$ \\
$\alpha$ & first Euler angle & $\mathcal{U}(0,\pi)$ \\
$\beta$ & second Euler angle & $\mathcal{U}(0,\pi/2)$ \\
$\gamma$ & third Euler angle & $\mathcal{U}(0,\pi)$ \\
$\log \Upsilon$ & stellar mass-to-light ratio & $\mathcal{U}(-2,2)$ \\
$\log \Omega_{\rm p}$ & bar pattern speed & $\mathcal{U}(0,2)$ \\
\hline
\end{tabular}
\end{table*}

\subsubsection{Model parameters and sampling}

In the model described above, there are 12 free parameters in the potential model, which are listed in Table~\ref{tab:model_parameters}. For the fiducial density profile for sampling the initial conditions of the orbital library (see Section~\ref{subsec:orbitlibrary}), we adopt $h_R=3.5~\rm kpc$ and $h_z=1.2~\rm kpc$ to ensure the spatial distribution of orbits completely covers the data extent. For the choice of the 3D density constraints, we segment the density model into 600 uniform cells in the cylindrical coordinate system, with $R$ ranging from $0~\rm to~10~\rm kpc$, $\phi$ ranging from $0~\rm to~2\pi$ and $z$ ranging from $-3~\rm to~3~kpc$. The choice of the 3D density grid is not optimal, as the grid should be denser in the central region of the galaxy, and it should also respect the azimuthal bisymmetry and $z$--reflection symmetry as the model is enforced to avoid repetitive binning. These will be accommodated in future versions of \textsc{SchwarMAX}. 

We sample the model parameters with a Markov chain Monte Carlo (MCMC) sampler implemented in \textsc{BlackJAX} \citep[][]{cabezas2024blackjax}, adopting uniform or log-uniform priors over the parameter ranges listed in Table~\ref{tab:model_parameters}. We run the MCMC sampler using 32 chains for 1500 steps for each chain. For reproducibility, we test the code performance using publicly available A100 GPU on the \texttt{Google Colab} platform. Each model $\chi^2$ evaluation takes $\approx1.3$~seconds, and the MCMC sampler runs for $\approx17-18$~hours.

\subsection{Data versus best-fit model}  \label{sec:mock_test_results}

We apply the model described in Section~\ref{sec:model} to this mock IFU data set in order to recover the potential parameters of the galaxy. The posterior distributions of the model parameters are shown in Fig.~\ref{fig:posterior_main}, where the red lines denote the ``ground truth'' density parameters. We define these ``ground truth'' parameters by fitting the parametrised density profiles to the three-dimensional stellar and dark matter particle distributions in the $N$-body snapshot. Because the adopted parametric forms are intrinsically different from the exact particle distributions in the simulation, these fitted values should be regarded as the best parametric representation of the true galaxy, subject to systematic uncertainty. Within this framework, 8 out of the 12 model parameters are recovered within or close to the $1\sigma$ uncertainties, including the bar pattern speed. The remaining four parameters are also recovered within $3\sigma$. 

The surface luminosity density, line-of-sight velocity, velocity dispersion, $h_3$, and $h_4$ maps of the best-fitting model are shown in the left column of Fig.~\ref{fig:data_vs_model}. The corresponding mock data are shown in the middle column, and the residuals normalised by the observational errors are shown on the right. The bulk photometric and kinematic structures of the galaxy are recovered well. The density map agrees well at low vertical height, but a relative mismatch of $\approx20\%$ appears at large vertical height. This mismatch arises from the $X\leftrightarrow-X,~Y\leftrightarrow-Y$ asymmetry present in the mock data from the $N$-body simulation, which reflects intrinsic non-axisymmetry at large galactic height. Since this feature is not included in our model, it cannot be reproduced. However, this is a systematic limitation of the adopted model family and does not significantly affect the recovery of the global model parameters.

\begin{figure}
    \centering
    \includegraphics[width=0.49\textwidth]{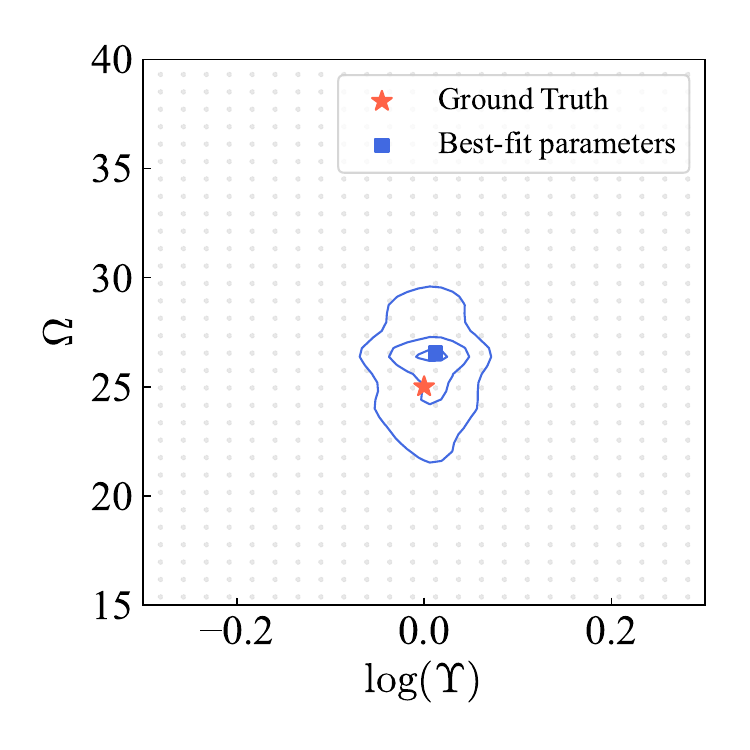}
    \caption{A grid search result for the best-fit mass-to-light ratio and bar pattern speed. The gravitation potential of the dark matter and stars are fixed to the ground truth in the simulation though azimuthal harmonic expansion. The blue contours shows the $1\sigma$, $3\sigma$, and $5\sigma$ contour. The solid blue square is the model with the highest likelihood, and the red star is the true mass-to-light ratio and pattern speed for reference. }
    \label{fig:grid_search_single}
\end{figure}

The recovered pattern speed is $25.35^{+1.5}_{-1.2}~\rm km/s/kpc$, compared to the reference value of $\approx 25~\rm km/s/kpc$. The pattern speed is therefore recovered with an accuracy of $\lesssim10\%$, consistent with previous Schwarzschild modelling results \citep{Vasiliev2020, Tahmasebzadeh2022, Tahmasebzadeh2024, Jin2025}. In contrast, the mass-to-light ratio, which has been recovered well in previous models, is underestimated by more than $1\sigma$ in our model. This may be caused by the mismatch between the adopted density-potential model and the true stellar distribution in the galaxy. To verify this, we perform an analysis similar to that of \citet{Vasiliev2020}, in which we simplify the model by fixing the stellar and dark matter density-potential profiles to the reference distributions in the simulation, leaving the stellar mass-to-light ratio and the bar pattern speed as the only free parameters. These reference density-potential profiles are obtained directly from the $N$-body snapshot, using the azimuthal-harmonic expansion (the \texttt{CylSpline} potential in \textsc{Agama}, \citealt{Vasiliev2019}). We then explore the model behaviour over a regular grid in bar pattern speed and mass-to-light ratio, as a grid search is more efficient than MCMC in two dimensions. The results of this simplified model are shown in Fig.~\ref{fig:grid_search_single}. The true pattern speed and mass-to-light ratio are indicated by the red star, the maximum-likelihood model is marked by the blue star, and the $1\sigma,~3\sigma,~5\sigma$ confidence contours are shown by the blue lines. In this simplified setup, the mass-to-light ratio is recovered correctly when the adopted density-potential profile matches that of the galaxy.

\subsection{Recovery of the density profile}

\begin{figure}
    \centering
    \includegraphics[width=0.49\textwidth]{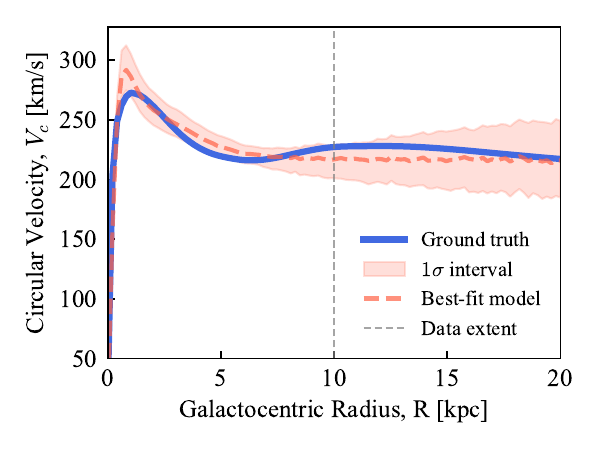}
    \caption{A comparsion between the recovered rotation curve of the galaxy with the best-fit potential model and the true rotation curve of the galaxy. The median and $1\sigma$ interval of the model is shown in the red dashed line and the shaded region, and the ground truth is shown in blue. The grey vertical dashed line denotes roughly the radial extent of the data.}
    \label{fig:rotation_curve}
\end{figure}

Using the fitted density parameters, we reconstruct the potential and density profile of the galaxy. In Fig.~\ref{fig:rotation_curve}, we compare the circular velocity curve of the axisymmetrised reconstructed potential with that of the $N$-body simulation used to generate the mock data. The blue line shows the reference rotation curve, while the red dashed line and shaded band show the median and $1\sigma$ interval of the reconstructed model. Within the radial extent covered by the mock data, the rotation curve is recovered with a maximum deviation of $10\%$. A similar deviation is also present in the model fitted directly to the actual particle distribution of the simulation, indicating that the mismatch is dominated by the systematic uncertainty of the adopted density parametrisation.

\begin{figure*}
    \centering
    \includegraphics[width=0.99\textwidth]{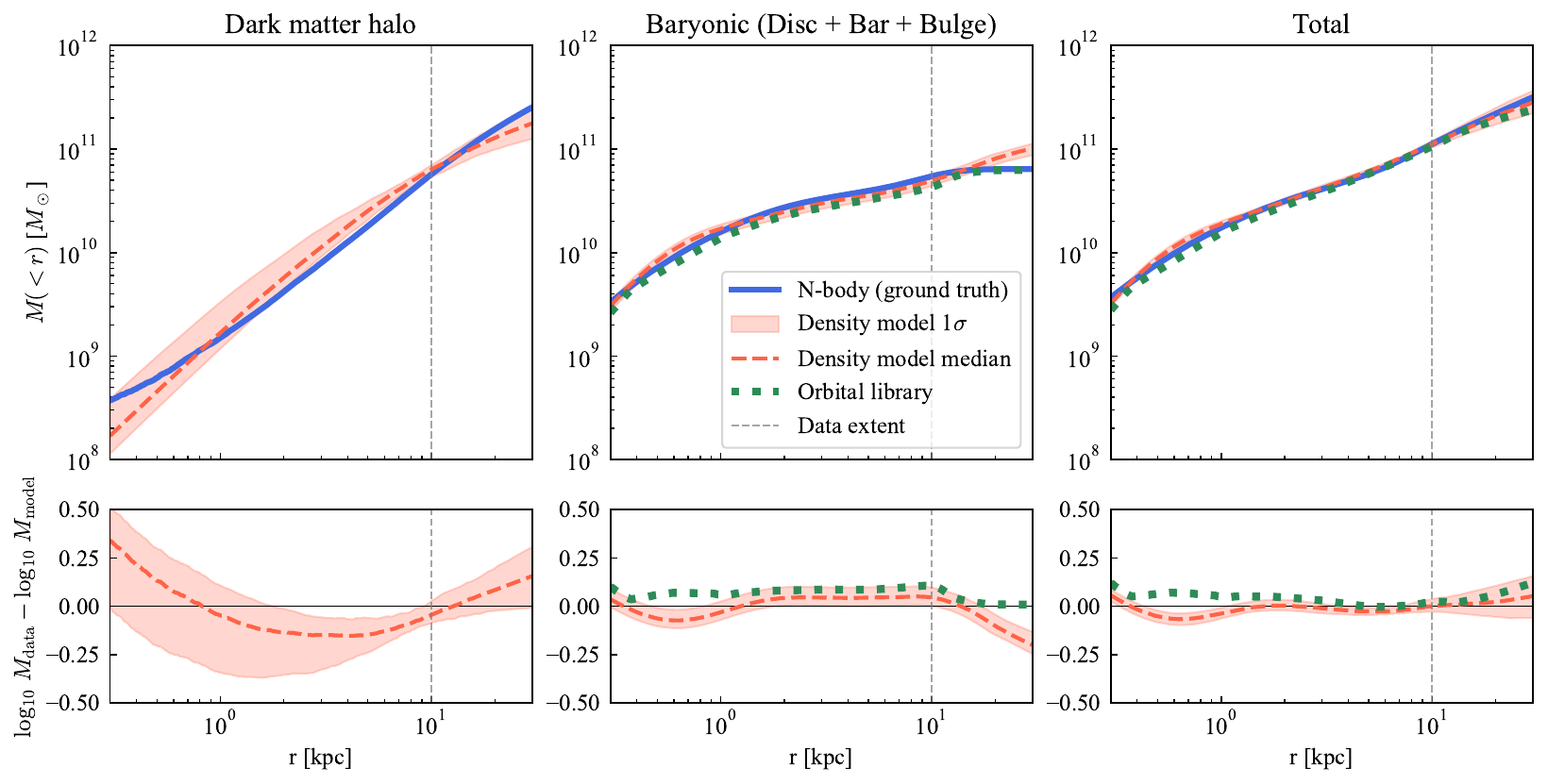}
    \caption{A comparison between the recovered mass profile of the dark matter/stars and the recovered model. {\it Left:} the enclosed mass profile of the dark matter. {\it Middle:} the enclosed mass profile of the stars. {\it Right:} the enclosed mass profile of both dark matter and stars. The top panels shows the enclosed mass as a function of radius, and the bottom panels show the difference between the recovered models and the true mass profile. The blue line represent the original galaxy, red dashed line and the shaded region represent the best-fit density model, and the green dashed line represent the orbital library, which is only for stars. }
    \label{fig:enclosed_mass}
\end{figure*}

We further compare the radial mass profiles of the model and the original galaxy. Fig.~\ref{fig:enclosed_mass} shows the enclosed mass of the dark matter, stellar, and total components in the left, middle, and right panels, respectively. We find good agreement between the model and the reference galaxy for $r\gtrsim1~\rm kpc$, with deviations of up to $\approx0.1~\rm dex$. At $r\lesssim1~\rm kpc$, the model shows a more substantial mismatch in the dark matter profile. This occurs because the dark matter halo in the simulation deviates from the assumed NFW form at these radii. However, since dark matter contributes only subdominantly to the total mass in the central region, this mismatch does not compromise the overall robustness of the model. Overall, the rotation curve and enclosed mass profile are both recovered well.

\subsection{Reconstructed orbital structure}

\begin{figure*}
    \centering
    \includegraphics[width=0.99\textwidth]{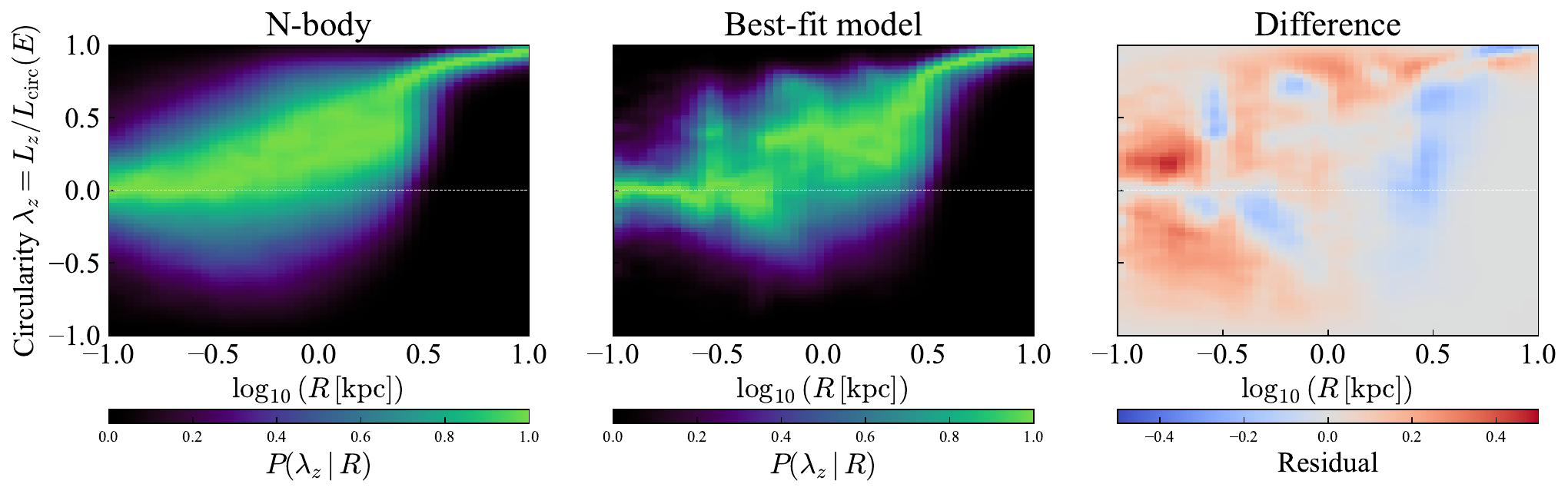}
    \caption{Reconstruction of the orbital structure of the galaxy demonstrated as the orbital circularity distribution as a function of radius. {\it Left:} the column-normalised distribution of orbital circularity of the original N-body simulation. {\it Middle:} the same for the best-fit orbital library. {\it Right:} the differences between the original galaxy and the best-fit orbital library. }
    \label{fig:orbits}
\end{figure*}

One of the main advantages of Schwarzschild modelling is its ability to recover the orbital structure of a galaxy \citep{Zhu2018}. We illustrate this in Fig.~\ref{fig:orbits}, using the same plane of radius and orbital circularity introduced by \citet{Zhu2018}. The orbital circularity, $\lambda_z$, is defined as the angular momentum $L_z$ normalised by the circular angular momentum at the same orbital energy, $L_{\rm circ}(E)$. In this definition, a prograde circular orbit has $\lambda_z=1$, a retrograde circular orbit has $\lambda_z=-1$, and a radial or polar orbit has $\lambda_z=0$. The left panel of Fig.~\ref{fig:orbits} shows the radial distribution of orbital circularity in the original galaxy, where the circularity distribution is column-normalised to remove the influence of the radial stellar density profile. To obtain the same quantity for our model, we interpolate and sample each orbit regularly in time. For each sampled point, we calculate both its circularity and galactocentric radius. The resulting orbit distribution is shown in the middle panel of Fig.~\ref{fig:orbits}. This effectively averages the circularity of each orbit, because the angular momentum is not strictly conserved in a barred potential. The right panel shows the difference between the original galaxy and our model.

At small radii, $\log R\lesssim0$, the bulge population dominates and exhibits little net rotation together with a broad spread in circularity. This dynamically hot component is reconstructed well in the orbital library, with similar rotational support and dispersion. Around the bar region, $0\lesssim \log R\lesssim0.5$, the stellar population shows strong prograde rotation, but only a small fraction of orbits have $\lambda_z\approx1$. This is expected, since the bar is dominated by elongated bar-supporting $x_1$ orbits rather than nearly circular disc orbits. The recovered orbital structure in this region is broadly consistent with that of the original galaxy, but the model does not fully recover the high-circularity population within the bar. This suggests that the bar-supporting orbit families are not reproduced perfectly, which might be because of the mismatch between the adopted analytic bar potential and the real one. Outside the bar region, $\log R\gtrsim0.5$, the disc rapidly becomes dominant, and circular orbits correspondingly dominate the circularity distribution. These are recovered correctly in our model.

\subsection{Model verification in more projection angles}

\begin{figure}
    \centering
    \includegraphics[width=0.5\textwidth]{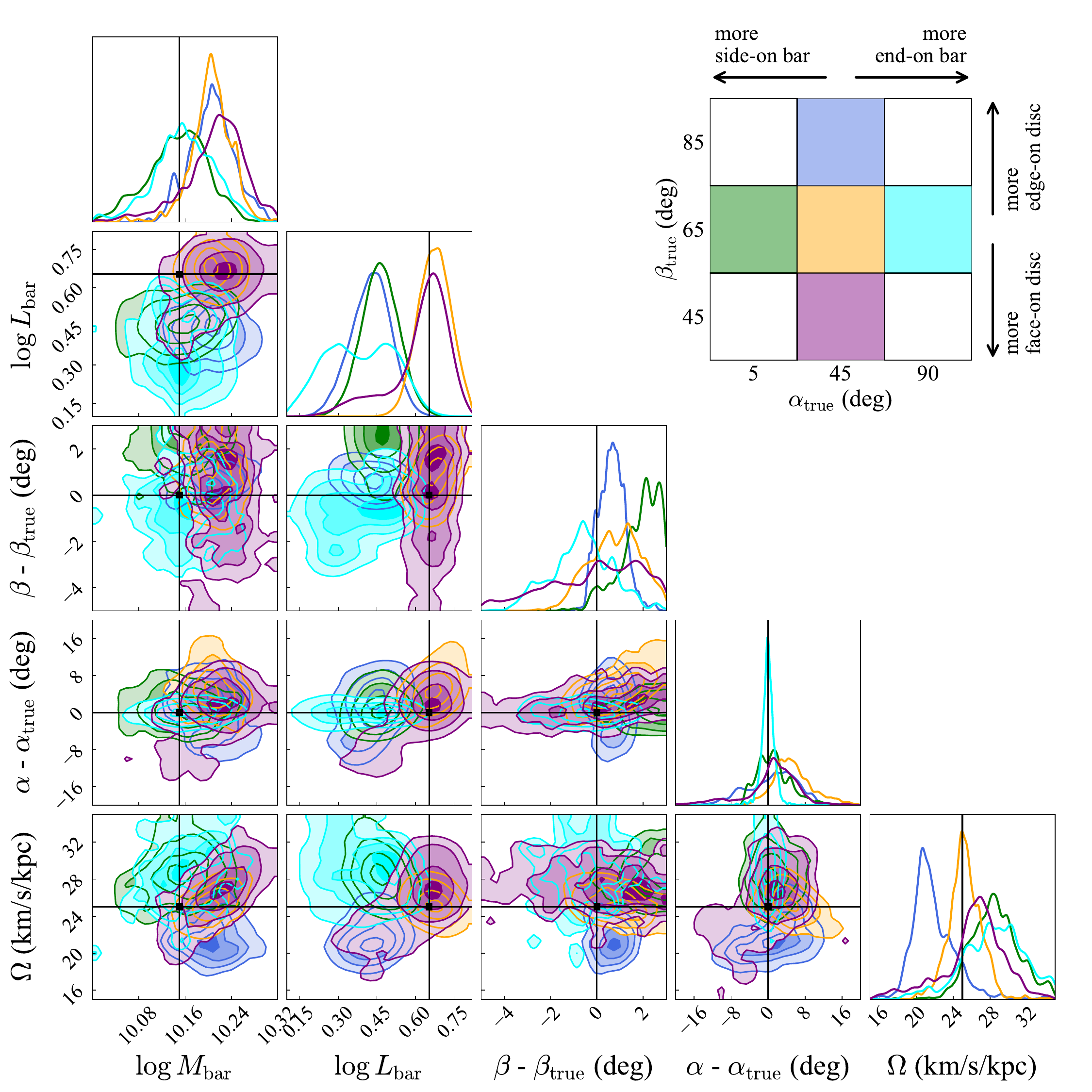}
    \caption{The stacked posterior of the bar mass, length, $\beta$, $\alpha$, and bar pattern speed of mock data with various projection angles. The mock data with $(\alpha, \beta)=(5^\circ, 65^\circ)$ is shown in green, $(\alpha, \beta)=(45^\circ, 65^\circ)$ in orange, $(\alpha, \beta)=(90^\circ, 65^\circ)$ is in aqua, $(\alpha, \beta)=(45^\circ, 85^\circ)$ in blue, and $(\alpha, \beta)=(45^\circ, 45^\circ)$ in purple.}
    \label{fig:posterior_varying_angles}
\end{figure}

To test the robustness of the model at other projection angles, we apply the same procedure to the same $N$-body galaxy, but generate mock data at different inclinations and in-plane bar orientations. Specifically, we construct four additional mock data sets with $(\alpha, \beta) = (5^\circ, 65^\circ)$, $(90^\circ, 65^\circ)$, $(45^\circ, 85^\circ)$, and $(45^\circ,~45^\circ)$. The mock data with $(5^\circ, 65^\circ)$ and $(90^\circ, 65^\circ)$ have the same inclination as the fiducial case, but the bar is viewed nearly side-on and end-on in the image plane, respectively. The cases $(45^\circ, 85^\circ)$ and $(45^\circ,~45^\circ)$ have the similar in-plane bar angle as the fiducial model, but correspond to a more edge-on view ($\rm inclination\approx85^\circ$) and a more face-on view ($\rm inclination\approx45^\circ$), respectively. Applying the same modelling procedure, we obtain posterior distributions for each mock observation analogous to those shown in Fig.~\ref{fig:posterior_main}.

In all cases, the halo and disc parameters are recovered as well as in the fiducial model. In Fig.~\ref{fig:posterior_varying_angles}, we show the posterior distributions of five selected parameters, $\log M_{\rm bar},~\log L_{\rm bar},~\beta,~\alpha$, and $\Omega_p$, all of which are closely associated with the bar. The posteriors for the four additional mock data sets are shown in blue for $(45^\circ, 85^\circ)$, aqua for $(90^\circ, 65^\circ)$, green for $(5^\circ, 65^\circ)$, and purple for $(45^\circ, 45^\circ)$. For comparison, the posterior of the fiducial case, $(45^\circ, 65^\circ)$, is also shown in orange. The lower-right panel gives the posterior distribution of the bar pattern speed. Our model recovers the bar pattern speed broadly consistent with the true value in all five mock observations across this range of inclinations and bar angles, although the blue cases show a slightly underestimated bar pattern speed. This could be because the projected bar carries less spatial and kinematic information when the disc is more edge-on. The bar length is underestimated in the fiducial mock data with an edge-on disc or the bar's major or minor axis is aligned to the image plane. By contrast, the bar length is recovered correctly when the galaxy is viewed at lower inclination and the bar major and minor axes subtend a larger angle on the image plane, implying that the bar parameters are easier to recover in these geometries.

\section{Discussion}

\subsection{Impact of different pattern speed on the kinematics}

\begin{figure*}
    \centering
    \includegraphics[width=0.99\textwidth]{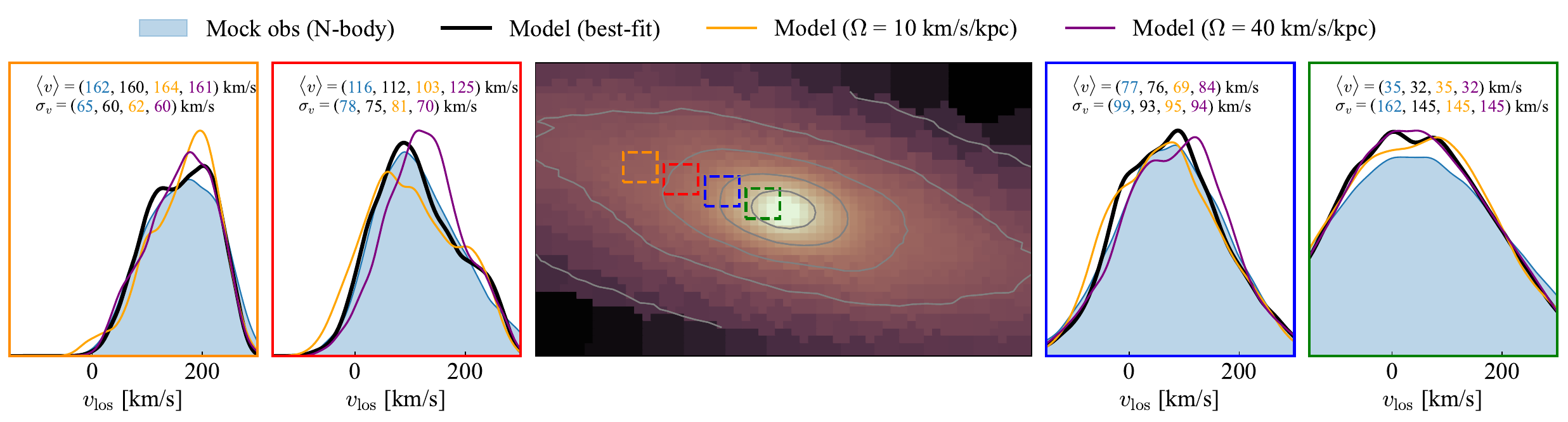}
    \caption{A comparison of the LOSVD distributions in different block of the galaxy and model LOSVD with different bar pattern speed. Four panels in the left and right of the middle surface density map are the LOSVD in the block with the corresponding colours. The original LOSVD of the galaxy in the block is shown in the blue shaded distribution. The model with the best-fit pattern speed is shown in the black line, and the model with low ($10~\rm km/s/kpc$) and high ($40~\rm km/s/kpc$) pattern speed are shown in orange and purple, respectively. The mean and dispersion of LOSVD is denoted in the texts in each panel with the corresponding colour, too.}
    \label{fig:vlos_vary_Omega}
\end{figure*}

The pattern speed of the galactic bar is one of the most important quantities to constrain in external barred galaxies. Although our model shows that the bar pattern speed can be recovered reliably, it is useful to identify the specific kinematic signatures that carry this information. After sampling the posterior for the mock data with $(\alpha,\beta) = (45^\circ,~65^\circ)$, we fix all density parameters to their best-fitting values and re-integrate the orbits, recalculating the orbital weights while varying the pattern speed from $10$ to $40$~km/s/kpc, corresponding to the true pattern speed $\pm15~\rm km/s/kpc$. We then compare the LOSVD along the galactic bar for models with different pattern speeds.

Fig.~\ref{fig:vlos_vary_Omega} shows the LOSVD in four regions along the bar major axis, from the bar centre to the bar end. The reference LOSVD is shown by the blue shaded distribution, while the models with pattern speeds of $10$, $25$ (best-fitting), and $40~\rm km/s/kpc$ are shown by the orange, black, and purple lines, respectively. In the block closest to the bar centre, shown in the rightmost panel, there is little difference among the three models, indicating that this region provides little leverage on the pattern speed. By contrast, in the two intermediate regions of the bar, shown by the blue and red blocks, the best-fitting model reproduces the reference LOSVD well. The two models with incorrect pattern speeds have broadly similar LOSVD shapes, but their mean velocities are offset because of the incorrect bar rotation rate. In the red block, the velocity dispersion also differs among the models. This indicates that the mean-velocity and velocity-dispersion maps in the central bar region already contain significant information on the pattern speed. At the bar end, shown by the orange block, the mean and dispersion of the LOSVD are nearly the same for all three models, but the shapes of the LOSVDs differ. The best-fitting model shows a flatter leading side, in closer agreement with the reference LOSVD, whereas the models with incorrect pattern speeds are noticeably peakier. This difference is reflected in the higher-order Gauss--Hermite moments, which therefore provide an additional constraint on the pattern speed in the outer bar region.

\subsection{Modelling with only mean and dispersion of LOSVD}

\begin{figure*}
    \centering
    \includegraphics[width=0.99\textwidth]{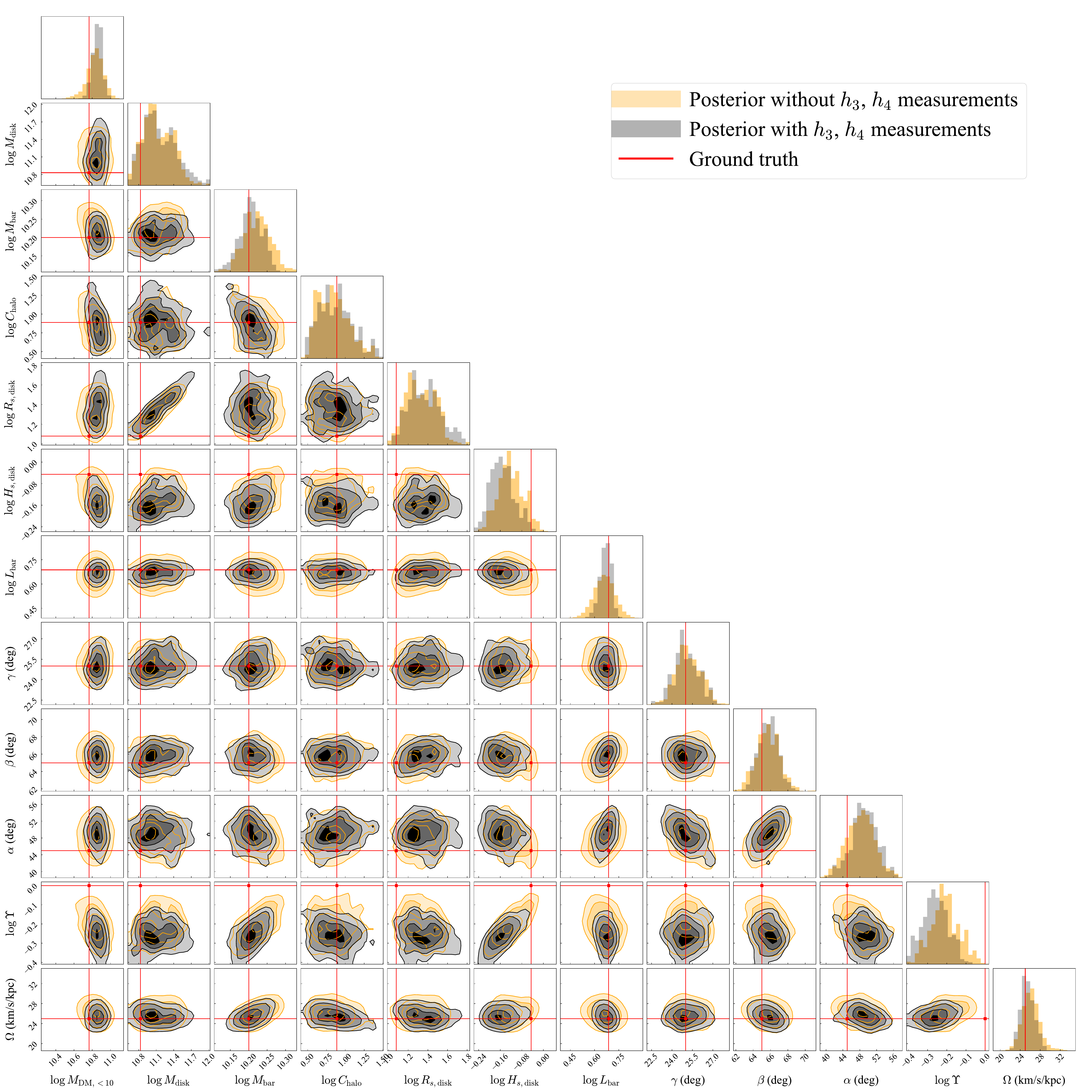}
    \caption{A comparison between the model fitted with and without $h_3$ and $h_4$ measurements. The posterior fitted without $h_3$ and $h_4$ is shown in orange, and the one with $h_3$ and $h_4$ is shown in black. The red line denotes the true value of each parameter. }
    \label{fig:posterior_noh3h4}
\end{figure*}

Since the mean and dispersion maps of the LOSVD already provide constraints on the pattern speed, we next test whether the Schwarzschild model can still recover the bar pattern speed without direct $h_3$ and $h_4$ measurements. We apply the same modelling and sampling procedure to the mock data with $(\alpha,\beta) = (45^\circ,65^\circ)$, but do not provide the measured $h_3$ and $h_4$ maps to the model. These moments are not removed entirely from the kinematic term $\chi_{\rm kine}^2$; instead, they are retained as a weak model prior. Since the values of $h_3$ and $h_4$ are typically small when $v$ and $s$ are chosen appropriately, we impose a Gaussian prior on $h_3$ and $h_4$ in each spatial aperture with mean $0$ and dispersion $0.15$, i.e. $h_3, h_4\sim\mathcal{N}(0,~0.15)$. 

The resulting posterior distribution for the model without direct $h_3$ and $h_4$ measurements is shown in Fig.~\ref{fig:posterior_noh3h4} by the orange contours and histograms, while the model fitted with the full $h_3$ and $h_4$ information is shown in black for comparison. Most model parameters are still recovered well, but with substantially broader posterior distributions than in the case where $h_3$ and $h_4$ are explicitly included. As expected from the discussion above, the bar pattern speed is still recovered correctly using only the $V_{\rm los}$ and $\sigma_v$ constraints, although the accuracy decreases modestly from $\approx7$--$8\%$ to $\approx11\%$. This is an encouraging result, because it suggests that the Schwarzschild model may still be applicable to barred galaxies observed with lower-quality kinematic data, such as those available in large IFU surveys, like MaNGA \citep{Bundy2015}.

\subsection{Future code developments}
\label{subsec::future_develop}
One important direction for future development is to make the full Schwarzschild pipeline more naturally compatible with gradient-based statistical inference. Although \textsc{SchwarMAX} is implemented in \textsc{JAX}, and thus in principle supports automatic differentiation, in practice we were not yet able to exploit this feature to perform posterior sampling with Hamiltonian Monte Carlo (HMC). The main difficulty is computational rather than conceptual. In the present implementation, each likelihood evaluation requires constructing a large orbital library with a long chain of orbital integration and performing NNLS iterations to determine the orbital weights. Reverse-mode differentiation through this entire calculation is prohibitively expensive in both memory and runtime.

\textbf{Orbital weights marginalisation with HMC:} A possible way forward is therefore to replace part of the orbital integration with a differentiable surrogate model. In particular, one could train an orbital emulator that maps orbital initial conditions and potential parameters directly to the corresponding projected density and kinematic contributions of each orbit. If such a surrogate can reproduce the orbital observables with sufficient accuracy, then constructing the orbital library would become much cheaper and, more importantly, differentiable end-to-end. This would make it possible to evaluate gradients of the model's likelihood with respect to the potential parameters at much lower cost, thereby enabling more efficient posterior exploration with HMC.

Such a development would also be interesting from a more statistical perspective to help address the distribution function marginalisation problem raised in \citet{Magorrian2006}. In most practical implementations, including the present one, the orbital weights are optimised at fixed potential to find one best-fit combination of the weights. If the orbit library can be evaluated rapidly and differentiated efficiently, it may become feasible to move beyond this approximation and explore models in which the orbital weights are treated more explicitly as model nuisance parameters during the optimisation, thereby effectively marginalising over the distribution functions. This remains a challenging high-dimensional problem, but modern differentiable surrogates and optimisation methods may provide a practical route that was previously out of reach. 

\textbf{Dust extinction:} Dust extinction is another important caveat that is not included in the present implementation. In Schwarzschild modelling, the observed surface-brightness distribution is used to constrain the underlying stellar density, so spatially varying dust extinction can bias the inferred luminosity distribution and, consequently, the deprojected mass model adopted for the orbit library. Dust can also affect the kinematic constraints, since the observed line-of-sight velocity distribution is luminosity-weighted and therefore depends on which stellar populations remain visible along a given line of sight. In barred disc galaxies, where dust lanes are often prominent, this may introduce local distortions in the measured velocity moments in addition to the photometric bias. Our mock tests do not include these effects, and therefore represent an idealized case. A more realistic application to dusty systems may require either extinction corrections, masking of heavily obscured regions. This will be tested in future studies.

\textbf{Deprojection:} We modelled the stellar density distribution in this work with Miyamoto-Nagai disc and bar models in \citet{Dehnen_Aly2023} as these densities have analytic potential counterparts. However, these analytic density-potential pairs cannot guarantee that they describe the 3D density and the 2D projected density of the galaxy well. For example, most of the galactic discs follow exponential profiles (including the fiducial N-body model we used here), which are different from the Miyamoto-Nagai disc. The density mismatch could introduce an internal inconsistency between the 2D surface density and the 3D density constraints we imposed in the model. However, as we demonstrated above, even if the chosen density model differs from the galaxy, the high flexibility of the Schwarzschild method still allows us to fit the data well and recover the model parameters and orbital structure of the galaxy correctly. It sacrifices the 3D density self-consistency constraints, which biases the stellar $M/L$ ratio. The model could achieve higher accuracy if one applies deprojection techniques to the 2D projected surface density before implementing the Schwarzschild method. Deprojection techniques have frequently been used in previous Schwarzschild models, either through MGE \citep{Emsellem1994} or parametric models \citep{Dattathri2024}. Once the galaxy is deprojected, the potential can be computed using azimuthal harmonic expansion as described in Section~\ref{subsec::density}. This could improve the model self-consistency as the 3D density is deprojected from the 2D projected density map. However, the evaluation of the potential model with azimuthal-harmonic expansion with sufficient resolution is typically twice slower than that when using analytic density-potential pairs. While both methods have their benefits, one could decide on the best approach based on the purpose.

\textbf{Schwarzschild modelling of the Milky Way:} Although this code is currently designed for external galaxy observation with only LOSVD available, it is interesting to extend the application of the Schwarzschild model to the Milky Way. The Milky Way is challenging to model accurately using Jeans or action-based methods due to the central stellar bar. Dynamical modelling of the Galactic bar has been done using the M2M model \citep[e.g.][]{Portail2017} and the Schwarzschild model \citep{Wang2013}, too. However, both were done mostly with integrated-light stellar kinematics. Recently, \citet{Clarke2022} compared the M2M model in \citet{Portail2017} to the discrete stellar kinematics from Gaia \citep{GaiaDR3} and VVV \citep{Minniti2010} to update the pattern speed of the Milky Way. The kinematics of the Galactic bar region have been well resolved by the long-period variables \citep{Grady2020, Hey2023, Zhang2024a, Zhang2024b}, but there is no update on the particle/orbital-based dynamical model of the Milky Way using resolved stellar kinematics. \citet{Khoperskov2025a, Khoperskov2025b} constructed a chemo-chrono-dynamical model of the Milky Way with a Schwarzschild-like orbit-superposition method but with a fixed gravitational potential. With the long-period variable sample in the upcoming Gaia DR4, a self-consistent chemodynamical model of the Milky Way could be worth exploring. 

\section{Conclusions}

In this work, we presented \textsc{SchwarMAX}, an orbit-superposition forward-modelling framework designed for barred galaxies and implemented with GPU-oriented parallelisation in \textsc{JAX} that is publicly available. Our model combines analytic density-potential pairs for the dark matter halo, stellar disc, and rotating stellar bar with a robust orbital integrator, NNLS-based orbital reweighting, and posterior sampling of the global model parameters. Relative to traditional Schwarzschild implementations, the main goal of \textsc{SchwarMAX} is to make the modelling more computationally practical while retaining the ability to infer both the underlying potential and the orbital structure. With \textsc{SchwarMAX}, we achieved an order-of-magnitude speedup over the previous code, enabling us to explore the parameter space in more than 10 dimensions using modern Bayesian inference methods.  

The method has four main ingredients. First, we construct the gravitational potential. Second, we generate an orbital library by sampling initial conditions and integrating the orbits in the rotating frame of the bar. Third, we project the orbits to the image plane and convert their contributions to the intrinsic density, surface luminosity density, and LOSVD represented by Gauss--Hermite coefficients. Finally, we determine the orbital weights using a regularised non-negative least-squares solver and sample the model potential parameters with MCMC. To approximately account for the uncertainty in orbital-weight determination, we further average over bootstrap realisations of the fitted orbital weights.

We verified the method using mock IFU observations generated from a barred $N$-body simulation. We constructed a dynamical model consisting of a dark matter halo, a stellar disc, and a rotating stellar bar with 12 free parameters in total. Most of the model parameters are retained with good accuracy. In particular, the bar pattern speed is recovered within $10\%$ accuracy. The model reproduces the projected density and kinematic maps well, recovers the rotation curve within $\sim10\%$ over the radial extent of the data, and recovers the radial mass profile of the galaxy to within $\sim0.1$ dex. The reconstructed orbital structure also broadly matches that of the original galaxy, including the dynamically hot central component, the strongly prograde bar region, and the outer disc-dominated circular orbits.

We also tested the robustness of the method across several viewing geometries. The dark matter halo, stellar disc, and bar pattern speed remain well recovered for all of the mock projections considered here. The bar parameters are most well recovered when the disc has moderate inclinations, and the bar major and minor axes are not perfectly aligned with the line-of-sight. We showed that the bar pattern speed is constrained primarily by the mean and dispersion of the LOSVD, and only secondarily by the higher-order Gauss-Hermite coefficients. We demonstrated that the bar pattern speed and other model parameters can be recovered well even without measurements of $h_3$ and $h_4$, although with reduced precision. 
%The bar parameters become less tightly constrained when the galaxy is more edge-on or when the stellar bar is closer to end-on, as expected from the reduced projected information content. Nevertheless, the pattern speed remains recoverable across all tested inclinations and bar orientations, which is one of the most encouraging results of this work. We further showed that the main kinematic constraints on the pattern speed arise from the mean velocity and velocity dispersion in the central bar region, together with higher-order LOSVD shape information near the bar end. Even when the observed $h_3$ and $h_4$ maps are removed, the pattern speed is still recovered, although with reduced precision.

Although we only demonstrated the performance of \textsc{SchwarMAX} on a mock barred galaxy, the code can be easily extended to apply to other stellar systems, such as dwarf and elliptical galaxies. With the demonstrated capability of \textsc{SchwarMAX}, the code is ready to be applied to modern IFU surveys such as MaNGA and GECKOS. 

%Overall, these tests show that \textsc{SchwarMAX} is already a practical Schwarzschild tool for modelling barred galaxies with IFU-like data. At the same time, its current limitations also point naturally to future developments, particularly in making the orbit-generation step more differentiable and statistically scalable. With further work, the same framework should be extendable beyond barred galaxies to other classical Schwarzschild applications, including dwarf galaxies and elliptical galaxies. We therefore view \textsc{SchwarMAX} both as a working barred-galaxy modelling code and as a step toward a new generation of Schwarzschild methods that better exploit modern accelerator hardware and statistical inference techniques.

% The last numbered section should briefly summarise what has been done, and describe
% the final conclusions which the authors draw from their work.
\section*{Data Availability}

\textsc{SchwarMAX} developed in this work is publicly available at \url{https://github.com/Hanyuan0908/SchwarMax}. 

\section*{Acknowledgements}

We thank the Surrey dynamical modelling workshop 2025 for inspiring this work. We thank the inspiring discussion with Jason Sanders, Daisuke Kawata, Ling Zhu, and Natsuki Funakoshi.

HZ thanks the Science and Technology Facilities Council (STFC) for a PhD studentship (grant number 2888170). VB and NWE acknowledge support from the Leverhulme Research Project Grant RPG-2021-205: "The Faint Universe Made Visible with Machine Learning". 
% The Acknowledgements section is not numbered. Here you can thank helpful
% colleagues, acknowledge funding agencies, telescopes and facilities used etc.

%%%%%%%%%%%%%%%%%%%%%%%%%%%%%%%%%%%%%%%%%%%%%%%%%%



\bibliographystyle{mnras}
\bibliography{bibliography} % if your bibtex file is called example.bib

%\begin{thebibliography}{99}
%\end{thebibliography}

%%%%%%%%%%%%%%%%%%%%%%%%%%%%%%%%%%%%%%%%%%%%%%%%%%

%%%%%%%%%%%%%%%%% APPENDICES %%%%%%%%%%%%%%%%%%%%%

\appendix

\section{Density-potential pairs of the galactic bar}
\label{Appendix:Dehnen_ALy_bar}

We adopt the analytic density-potential pair described in \citet{Dehnen_Aly2023}. The bar potential is constructed by convolving the variants of the Miyamoto-Nagai disc with a needle function. They introduced two variant families of the Miyamoto-Nagai disc, named $T$ and $V$ family, and we choose the $T_3$ and $V_4$ potential model. The parent axisymmetric potential-density pairs of these two potentials are
\begin{equation}
\begin{aligned}
\Phi_{T_3} &= -GM_b\left[
\frac{1}{\mathcal R}
+
\frac{a\left(Z-\frac{a}{3}\right)}{\mathcal R^3}
+
\frac{a^2 Z^2}{\mathcal R^5}
\right], \\
\rho_{T_3} &= \frac{M_b b^2}{4\pi} \left[
\frac{3}{\mathcal R^5}
+
\frac{5aZ^2\left(3\zeta^2-2a\zeta+a^2\right)}{\mathcal R^7\zeta^3}
+
\frac{35a^2Z^4}{\mathcal R^9\zeta^2}
\right].
\end{aligned}
\end{equation}
and
\begin{equation}
\begin{aligned}
\Phi_{V_4} = -GM_b\Bigg[
&\frac{1}{\mathcal R}
+ \frac{a\left(Z-\frac{2a}{5}\right)+\frac{b^2}{2}}
{\mathcal R^3} \\
&+ \frac{
3a\left[
\frac{1}{5}aZ(2Z-a)
+\frac{b^2}{2}\left(Z-\frac{a}{5}\right)
\right]}
{\mathcal R^5} \\
&+ \frac{a^2Z^2(aZ+3b^2)}
{\mathcal R^7}
+ \frac{7a^3b^2Z^4}
{2\mathcal R^9\zeta}
\Bigg], \\
\rho_{V_4} = \frac{3M_b b^4}{8\pi}\Bigg[
&\frac{5}{\mathcal R^7}
+ \frac{
7aZ^2\left(
5\zeta^4-4a\zeta^3+3a^2\zeta^2-2a^3\zeta+a^4
\right)}
{\mathcal R^9\zeta^5} \\
&+ \frac{
63a^2Z^4\left(2\zeta^2-2a\zeta+a^2\right)}
{\mathcal R^{11}\zeta^4}
+ \frac{231a^3Z^6}
{\mathcal R^{13}\zeta^3}
\Bigg].
\end{aligned}
\end{equation}
Replacing $\mathcal{R}^{-n}$ with $I_n$ as in Eq.~\ref{eqn:Rn_to_In} yields the barred potential-density pairs of $T_3$ and $V_4$.

%%%%%%%%%%%%%%%%%%%%%%%%%%%%%%%%%%%%%%%%%%%%%%%%%%

\section{Comparison with \textsc{Forstand}}  \label{Appendix:Forstand}

In this section, we compare \textsc{SchwarMAX} with another Schwarzschild modelling code, \textsc{Forstand} \citep{Vasiliev2020}. Although the orbit-superposition method has been used for galactic modelling for three decades, there is no published study that would directly compare two or more implementations of the method, either on mock data or on real galaxies. Here we do not aim at a comprehensive comparison, but only highlight the key differences between the codes:
\begin{itemize}
\item \textsc{Forstand} is built on top of \textsc{Agama} framework for galactic dynamics \citep{Vasiliev2019}. Although highly optimised and parallelised, it is a CPU-based code, and cannot compete in speed with GPU-native \textsc{SchwarMAX}. The construction of a $N_{\rm orb}=10^4$ library of orbits integrated for 100 dynamical times with an eight-order Runge--Kutta ODE solver takes $\mathcal O(10)$ seconds on a 16-core CPU, and the solution of a single optimisation problem with $10^4$ orbits and $3\,500$ constraints takes another $\mathcal O(10)$ seconds. On the other hand, \textsc{Agama} provides a much broader choice of analytic gravitational potential--density pairs and general-purpose Poisson solvers for arbitrary density profiles.
\item \textsc{Forstand} stores orbit kinematics on a 3d grid in $x', y', v_z'$ in the form of B-splines, and converts them to Gauss--Hermite coefficients before comparing with the observational data. Thus one can reuse the same orbit library for many different values of $\Upsilon$, rescaling the velocity axis by $\sqrt{\Upsilon}$ before converting B-splines into Gauss--Hermite coefficients; here $\Upsilon$ scales all mass components in the same proportion, rather than just stars. The code finds the value of $\Upsilon$ producing the lowest $\chi^2$ for the given orbit library by a simple grid search with fixed logarithmic increments/decrements in $\Upsilon$, before switching to building another orbit library for different values of potential parameters or $\Omega$. By contrast, \textsc{SchwarMAX} does not treat $\Upsilon$ differently from other model parameters, and constructs a new orbit library each time, since the orbit kinematics are directly converted into Gauss--Hermite coefficients on the fly.
\item All previous implementations of the Schwarzschild method treat the surface density profile as fixed (derived from photometry prior to dynamical modelling). It is then deprojected, i.e.\ converted to 3d density profile and then to the potential used for the orbit integration. This usually relies on the fact that profiles with ellipsoidal equidensity contours also look ellipsoidal in projection, so for a given observed flattening of the surface density profile and an assumed orientation of the model, one can uniquely determine the axis ratios of the 3d density (see e.g.\ section~3 in \citealt{vandenBosch2008} and section~3 in \citealt{Quenneville2022} for details). However, real galaxies, in particular barred ones, may not be well described by ellipsoidally stratified density profiles, or even combinations of several such profiles (such as the widely used multi-Gaussian expansion, \citealt{Emsellem1994}). 
Although it may be possible to use negative-weight components to produce more boxy models, it becomes difficult to ensure that the 3d density remains non-negative everywhere. An alternative, forward-modelling approach is to start from a parameterised, physically motivated 3d density profile and fit it to the surface density map under different choices of orientation angles \citep{deNicola2020,Dattathri2024}. In either case, the orbit-superposition model tries to fit the discretised 3d and 2d density constraints of this input profile, rather than the observational photometric maps directly. By contrast, \textsc{SchwarMAX} does not rely on a separate photometric modelling step, and builds a full dynamical model for every choice of parameters, fitting the actual surface density rather than its parametric approximation. This effectively extends the forward modelling approach from photometry to dynamics, but has a downside that if the adopted parametric density profile cannot provide a good fit to the actual surface density map, then the intrinsic (3d density) constraints coming from the parametric model and projected (surface density) constraints fitted to observations are in conflict and cannot be simultaneously satisfied to high precision.
\end{itemize}

\begin{figure}
\includegraphics{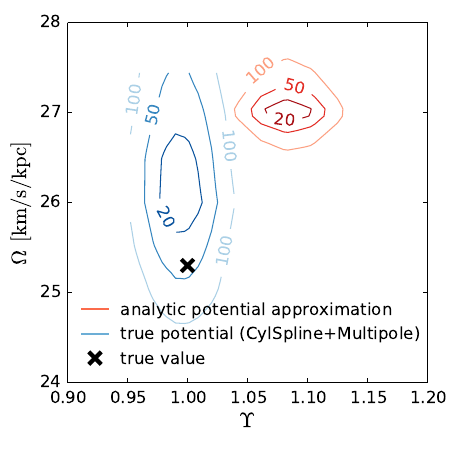}
\caption{Likelihood surface in the space of two model parameters (pattern speed $\Omega$ and $M/L$ ratio $\Upsilon$) for \textsc{Forstand} models run with the ``true'' density profile of both disc and halo (represented by \texttt{CylSpline} and \texttt{Multipole} potentials, respectively; blue contours) and with the approximate potential constructed from three analytic components, as described in Section~\ref{sec:mock_test_model_setup} (red contours). Shown are contours of $\chi^2_\mathrm{model} - \mathrm{min}\chi^2_\mathrm{model}$ averaged over 10 realisations of random noise in the mock kinematic maps. The true values of $\Omega$ and $\Upsilon$ are indicated by the black cross.}
\label{fig:chi2_forstand}
\end{figure}

There are also many other differences in implementation details between \textsc{Forstand} and \textsc{SchwarMAX}, e.g.\ the discretisation schemes for the 3d density profile and typical adopted tolerance on satisfying these constraints, creation of inital conditions for the orbit library, ODE integrators, optimisation problem solvers, etc. As such, a direct comparison of $\chi^2$ values produced by the two codes is challenging. Nevertheless, we ran two series of \textsc{Forstand} model fits using the same input kinematic data and two choices for the density model: one derived directly from the $N$-body snapshot and represented by the \texttt{CylSpline} expansion for the stars and \texttt{Multipole} for the dark halo (as in section~3.3 of \citealt{Vasiliev2020}), and the other using analytic potential--density pairs described in Section~\ref{sec:mock_test_model_setup}. We used $10^4$ orbits, four Gauss--Hermite moments plus the surface density constraints with a 1\% tolerance, and discretised the 3d density into $\sim 600$ spatial bins, adopting a relative tolerance of 10\% for these constraints. All model parameters except $\Omega$ and $\Upsilon$ were kept fixed to their true values (in the case of analytic density profiles, to those found by fitting the density profiles directly to the particle positions, although these fits are not perfect).
Figure~\ref{fig:chi2_forstand} shows the $\chi^2$ contours in the space of these two free parameters. Models with the true density profile fit the data better and recover the $M/L$ ratio perfectly, while overestimating the pattern speed by a few percent, while models with analytic profiles are somewhat worse and biased by $\lesssim 10\%$ in both parameters. Due to the scatter in $\chi^2$ values caused by chaotic orbits, as well as their dependence on other model parameters, such as the number of orbits or the 3d discretisation grid, it is not meaningful to compare the absolute values of $\chi^2$ or even the width of the contours between the two codes, but the inference from \textsc{Forstand} is broadly consistent with the results from \textsc{SchwarMAX} described in Section~\ref{sec:mock_test_results}.

% Don't change these lines
\bsp	% typesetting comment
\label{lastpage}
\end{document}